\documentclass{aa}

\usepackage{graphicx}
\usepackage{txfonts}

\begin{document}

\title{A radio-jet driven outflow in the Seyfert 2 galaxy NGC 2110?}

\author{L. Peralta de Arriba\inst{1},
        A. Alonso-Herrero\inst{1},
        S. Garc\'{\i}a-Burillo\inst{2},
        I. Garc\'{\i}a-Bernete\inst{3},
        M. Villar-Mart\'{\i}n\inst{4},
        B.~Garc\'{\i}a-Lorenzo\inst{5, 6},
        R. Davies\inst{7},
        D. J. Rosario\inst{8},
        S. F. H\"onig\inst{9},
        N. A. Levenson\inst{10},
        C. Packham\inst{11, 12},
        C.~Ramos~Almeida\inst{5, 6},
        M. Pereira-Santaella\inst{13},
        A. Audibert\inst{5, 6},
        E. Bellocchi\inst{14, 15},
        E. K. S. Hicks \inst{16},
        A. Labiano\inst{17},
        C.~Ricci\inst{18, 19}
        \and
        D. Rigopoulou\inst{3, 20}}
\authorrunning{L. Peralta de Arriba et al.}

\institute{Centro de Astrobiolog\'{\i}a (CAB), CSIC-INTA, Camino Bajo del
           Castillo s/n, 28692 Villanueva de la Ca\~nada, Madrid, Spain\\
           \email{lperalta@cab.inta-csic.es}
             \and
           Observatorio de Madrid, OAN-IGN, Alfonso XII, 3, E-28014 Madrid,
           Spain
             \and
           Department of Physics, University of Oxford, Keble Road, Oxford OX1
           3RH, UK
             \and
           Centro de Astrobiolog\'{\i}a (CAB), CSIC-INTA, Carretera de Ajalvir
           km 4, 28850, Torrej\'on de Ardoz, Madrid, Spain
             \and
           Instituto de Astrof\'{\i}sica de Canarias (IAC), Calle V\'{\i}a
           L\'actea, s/n, 38205 La Laguna, Tenerife, Spain
             \and
           Departamento de Astrof\'{\i}sica, Universidad de La Laguna, 38206 La
           Laguna, Tenerife, Spain
             \and
           Max-Planck-Institut f\"ur Extraterrestrische Physik, Postfach 1312,
           85741 Garching, Germany
             \and
           School of Mathematics, Statistics and Physics, Newcastle University,
           Newcastle upon Tyne NE1 7RU, UK
             \and
           School of Physics \& Astronomy, University of Southampton,
           Southampton SO17 1BJ, Hampshire, UK
             \and
           Space Telescope Science Institute, Baltimore, MD 21218, USA
             \and
           The University of Texas at San Antonio, One UTSA Circle, San Antonio,
           TX 78249, USA
             \and
           National Astronomical Observatory of Japan, National Institutes of
           Natural Sciences (NINS), 2-21-1 Osawa, Mitaka, Tokyo 181-8588, Japan
           \and
           Instituto de F\'{\i}sica Fundamental, CSIC, Calle Serrano 123, 28006 Madrid, Spain
           \and
           Departamento de F\'{\i}sica de la Tierra y Astrof\'{\i}sica, Fac. de
           CC F\'{\i}sicas, Universidad Complutense de Madrid, 28040 Madrid,
           Spain
             \and
           Instituto de F\'{\i}sica de Part\'{\i}culas y del Cosmos IPARCOS,
           Fac. CC F\'{\i}sicas, Universidad Complutense de Madrid, 28040
           Madrid, Spain
             \and
           Department of Physics \& Astronomy, University of Alaska Anchorage,
           Anchorage, AK 99508-4664, USA
             \and
           Telespazio UK for the European Space Agency, ESAC, Camino Bajo del
           Castillo s/n, 28692 Villanueva de la Ca\~nada, Spain
             \and
           N\'ucleo de Astronom\'ia de la Facultad de Ingenier\'ia, Universidad
           Diego Portales, Av. Ej\'ercito Libertador 441, Santiago, Chile
             \and
           Kavli Institute for Astronomy and Astrophysics, Peking University,
           Beijing 100871, China
           \and
         School of Sciences, European University Cyprus, Diogenes
         Street, Engomi, 1516 Nicosia, Cyprus} 

\date{Received Month DD, YYYY; accepted Month DD, YYYY} 

\abstract{We present a spatially-resolved study of the ionised gas in the
  central 2\,kpc of the Seyfert 2 galaxy NGC~2110 and investigate the role of
  its moderate luminosity radio jet (kinetic radio power of $P_\mathrm{jet} =
  2.3 \times 10^{43} \ \mathrm{erg\,s^{-1}}$). We use new optical integral-field
  observations taken with the MEGARA spectrograph at the Gran Telescopio
  Canarias, covering the $4300- 5200\,\AA$ and $6100-7300\,\AA$ ranges with a
  spectral resolution of $R\simeq 5000-5900$. We fit the emission lines with a
  maximum of two Gaussian components, except at the AGN position
    where we used three. Aided by existing stellar kinematics, we
  use the observed velocity and velocity dispersion ($\sigma$) of the emission
  lines to classify the different kinematic components. The disc component is
  characterised by lines with $\sigma \simeq 60-200\,\mathrm{km\, s}^{-1}$. The
  outflow component has typical values of $\sigma \simeq 700\,\mathrm{km\,
    s}^{-1}$ and is confined to the central $2.5\arcsec \simeq 400\,$pc, which
  is coincident with linear part of the radio jet detected in NGC~2110. At the
  AGN position, the [\ion{O}{III}]$\lambda$5007 line shows high velocity
  components reaching at least $1000\, \mathrm{km\,s}^{-1}$. This and the high
  velocity dispersions indicate the presence of outflowing gas outside the
  galaxy plane. Spatially-resolved diagnostic diagrams reveal mostly low
  ionisation (nuclear) emitting region (LI(N)ER)-like excitation in the outflow
  and some regions in the disc, which could be due to the
    presence of shocks.
  However, there is also Seyfert-like excitation
  beyond the bending of the radio jet, probably tracing the edge of the
  ionisation cone that intercepts with the disc of the galaxy. NGC~2110 follows
  well the  observational trends between the outflow properties and the jet radio power
  found for a few nearby Seyfert galaxies. All these pieces of information
  suggest that part of observed ionised outflow in NGC~2110 might
    be driven by the radio jet.  However, the
  radio jet was bent at radial distances of $\sim 200\,$pc (in projection) from
  the AGN, and beyond there, most of the gas in the galaxy disc is rotating.}


\maketitle


\begin{figure*}[h]

  \vspace{-1cm}
  \hspace{-4cm}
  \includegraphics[width=26cm]{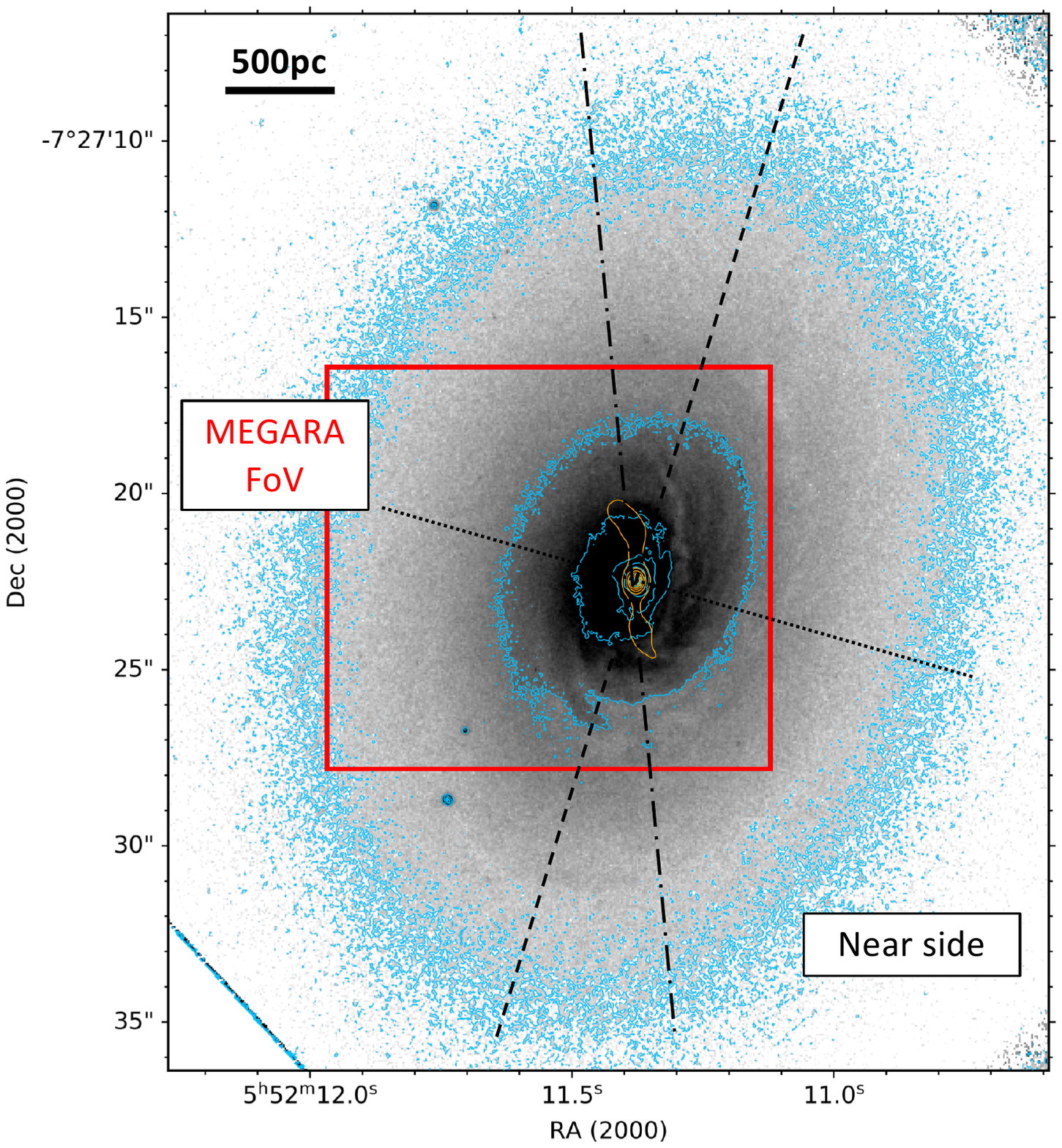}
\vspace{-2.5cm}
  \caption{\emph{HST}/WFPC2 optical image of NGC~2110 obtained 
     with the F791W  filter, shown on a linear scale (grey colors and blue
     contours). The orange contours (also on a linear scale) are the VLA radio 6\,cm emission from 
\cite{Ulvestad1983}. We also mark with a red rectangle the approximate size of the
    GTC/MEGARA FoV of $12.5\arcsec \times 11.3\arcsec$. The dashed
      and dotted lines are the position angle of
    the galaxy major axis \citep[PA $\simeq 163^\circ$ measured from the north
      to the east, see][]{Ferruit2004} and the minor axis,
      respectively. The dashed-dotted line is the PA of the slit used
      by \cite{GonzalezDelgado2002} and 
         discussed in Sect.~\ref{subsec:compstellarkins}.}
  \label{fig:photo}
\end{figure*}

\section{Introduction} \label{sec:intro}

Active galactic nuclei (AGN) are crucial for understanding galaxy evolution. AGN
feedback has been proposed as one mechanism to reproduce the observed number of
the most massive galaxies, because in its absence star formation would have been
too efficient \citep[e.g.,][]{Silk2012}. In particular, in moderate luminosity
AGN, the relationship between radiation driven outflows and radio jets remains
as an open problem. From the theoretical point of view, it is expected that
galaxies hosting radio jets could drive outflowing material
\citep[e.g.,][]{Weinberger2017a,Mukherjee2018,Talbot2022,Meenakshi2022}. From
the observational side, the capacity of high power radio jets to produce
energetic outflows is well demonstrated
\citep[e.g.,][]{Nesvadba2008,Vayner2017}.

The impact of low to moderate luminosity radio jets on the interstellar medium
(ISM) of their host galaxies has been studied mostly in local Seyfert galaxies,
low-luminosity AGN, and quasars. It has been shown that in the presence of a
good geometrical coupling, these jets can transfer successfully mechanical
energy to the ISM of their host galaxies
\citep[e.g.,][]{Combes2013,GarciaBurillo2014,Morganti2015,GarciaBernete2021,
  Maccagni2021,PereiraSantaella2022,Cazzoli2022, Speranza2022,
  Audibert2023}. Using a small sample of 
Seyfert galaxies, \citet{Venturi2021} detected enhanced widths of optical lines
in regions perpendicular to the radio jet, while Seyfert-like excited conical
shapes were aligned with the radio jet. They interpreted this as the result of
the interaction between the radio jet and outflowing material. In this context,
they proposed scaling relationships between both the ionised gas mass and
kinetic energy in the regions with enhanced line width and the radio jet power.

NGC~2110 is an early type spiral galaxy (SAB0) hosting a Seyfert 2 nucleus
\citep{McClintock1979}, observed at an intermediate inclination $i=42-65^\circ$
\citep[see e.g.,][and
  Fig.~\ref{fig:photo}]{Wilson1985,GonzalezDelgado2002,Kawamuro2020}. It
contains a radio jet, whose `S'-shaped jet was proposed to be the
result of its bending by the ram pressure of the host galaxy rotating gas
\citep{Ulvestad1983}. The jet appears brighter on the north side, which allowed
\cite{Pringle1999} to estimate that it is at an intermediate orientation with
respect to the plane of the galaxy. The luminosity of this jet lies at the high
end of the 20\,cm radio luminosity distribution of the early-type Seyfert
galaxies studied by \cite{Nagar1999}.

Several works studied NGC~2110 in detail using optical integral field unit (IFU)
spectroscopy \citep{GonzalezDelgado2002,Ferruit2004,SchnorrMueller2014} with
spectral resolutions $R=\lambda$/$\Delta \lambda \simeq 1000-2000$. They all
detected the presence of complex non-circular motions in the ionised gas, which
were interpreted as due to the presence of an outflow, inflow, and/or a minor
merger. Moreover, \citet{Rosario2010} demonstrated that the radio jet has
sufficient ram pressure to drive the emission-line outflow. Finally,
using Atamaca Large Millimeter Array (ALMA)
CO(2-1) observations, \citet{Rosario2019} discovered a nuclear region with faint
cold molecular gas emission (termed ``lacuna''), which is oriented in the
approximate direction of the radio jet and filled with warm molecular hydrogen
emission. These authors suggested that the radio jet and/or an AGN wind could be
responsible for this phenomenon.

\begin{table*}
  \caption{Parameters employed for the construction of cubelets and their line
    fits.}
  \label{tab:cubelets}
  \centering
  \begin{tabular}{c c c c c}
    \hline\hline
    Cubelet name &
    Rest-frame wavelength range & Rest-frame continuum regions &
    AoN & $A_\mathrm{min}$\\
    &
    (\AA) & (\AA) &
    & (10$^{-18}$ erg s$^{-1}$ cm$^{-2}$ \AA$^{-1}$)\\
    \hline
    H$\beta$ &
    4800--4920 & 4800--4820, 4900--4920 &
    1.5 & 4.0\\
    {[\ion{O}{III}]}$\lambda\lambda$4959, 5007 &
    4900--5065 & 4900--4920, 5045--5065 &
    2.0 & 4.0\\
    {[\ion{O}{I}]}$\lambda$6300 &
    6255--6345 & 6255--6275, 6325--6345 &
    1.5 & 3.0\\
    H$\alpha$ + [\ion{N}{II}]$\lambda\lambda$6548, 6583 &
    6475--6640 & 6475--6495, 6620--6640 &
    2.0 & 7.0\\
    {[\ion{S}{II}]}$\lambda\lambda$6716, 6731 &
    6660--6780 & 6660--6680, 6760--6780 &
    2.0 & 4.0\\
    \hline
  \end{tabular}
  \tablefoot{AoN is the minimum amplitude-over-noise used for fitting each
    emission lines and $A_\mathrm{min}$ is the minimum amplitude used for the
    detection of each emission line (see text for more details).}
\end{table*}

As part of the Galactic Activity, Torus, and Outflow Survey (GATOS), we obtained
new optical IFU observations of several Seyfert galaxies,
  including NGC~2110 
    using the Multi-Espectr\'ografo en GTC 
de Alta Resoluci\'on para Astronom\'{\i}a
\citep[MEGARA,][]{GildePaz2016,Carrasco2018} at the 10.4-m Gran Telescopio
Canarias (GTC) with $R=5000-5900$. Our observations 
cover the $\sim$4300--5200~$\AA$ and $\sim$6100--7300~$\AA$ spectral ranges, which allow
us to spatially resolve both the gas kinematics and the excitation mechanisms in
 NGC~2110. By revisiting this well-studied target, we plan to  use it as a
  benchmark for other forthcoming analyses of MEGARA observations within the
  GATOS survey.

This paper is structured as follows. Section~\ref{sec:data} describes the data
employed in this work. Section~\ref{sec:analysis} explains the analysis, the
classification of gas components, and the construction of the two-dimensional
maps. We discuss the properties of the disc and outflow regions of NGC~2110 in
Sects.~\ref{sec:disccomponent} and ~\ref{sec:outflowcomponent},
respectively. Finally, the summary and conclusions are presented in
Sect.~\ref{sec:conclusions}. We assume a systemic velocity of
$2335\pm 20\,$km s$^{-1}$, based on measurements of stellar absorptions from the
magnesium $b$ triplet at $\sim$5175 \AA\ by \citet{Nelson1995}. We adopt a
$\Lambda$ cold dark matter cosmology with $H_0$=70 km s$^{-1}$ Mpc$^{-1}$,
$\Omega_\mathrm{m}$=0.3, and $\Omega_\Lambda$=0.7, which leads to a luminosity
distance of $D_L= 33.6\,$Mpc and a physical scale of
$160.2\,\mathrm{pc}/\arcsec$.


\section{Observations} \label{sec:data}


\subsection{Optical IFU data from MEGARA} \label{subsec:optical_data}

We observed the central region of NGC~2110  (Program
  GTC27-19B, PI: Alonso Herrero) using the large compact
  bundle (LCB) of fibers of MEGARA (i.e.,  the IFU mode of this instrument).
This IFU provides a field of view (FoV) of $12.5\arcsec \times 11.3\arcsec$ (see
Fig.~\ref{fig:photo}), using 567 fibers of $0.62\arcsec$ in diameter. The
physical region covered for NGC~2110 by our MEGARA observations is 2.0 $\times$
1.8 kpc$^2$. We used two volume-phased holographic gratings. The LR-B grating
covers the spectral range $\sim$4300--5200~$\AA$ with $R\sim$5000. H$\beta$ and
the [\ion{O}{III}]$\lambda\lambda$4959, 5007 doublet are within this range for
NGC~2110. The LR-R grating covers the $\sim$6100--7300~$\AA$ range with $R$
$\sim$ 5900, thus including [\ion{O}{I}]$\lambda$6300, H$\alpha$,
[\ion{N}{II}]$\lambda\lambda$6548, 6583 and [\ion{S}{II}]$\lambda\lambda$6716,
6731. The on-source integration times were $\sim$1 hour (7 exposures of 520\,s
each) for LR-B and half hour (5 exposures of 360\,s each) for LR-R.

The observations were taken in two consecutive observation blocks on the night
of 9$^\mathrm{th}$ February 2021. The observing conditions were a dark
photometric sky and a seeing of $0.9\arcsec$. The position angle (PA) of the
observations was $-0.598\degr$, which is practically equivalent to the regular
orientation north up, east to the left.

The data reduction was performed using the official MEGARA pipeline
\citep{Pascual2021}. The data reduction process of this software includes bias
subtraction, bad-pixels masking, tracing fibres, wavelength calibration
(including versions corrected and uncorrected of barycentric velocity),
flat-field correction, flux calibration, sky subtraction and cube
construction. The flat-field images were taken with a continuum halogen lamp and
close in time to the target exposures. This allows us to use them for tracing
fibres without dealing with their known shifts that depend on the ambient
temperature. Also just before the target exposures, we obtained images
illuminated with a Thorium-Argon hollow cathode lamp, which were used for the
wavelength calibration. The spectrophotometric standard star HR~1544/HR~4554 was
used for the flux calibration of the observations with the LR-B/LR-R
grating. The final data cubes were produced with $0.3\arcsec$ spaxels, which is
the recommended size by the pipeline developers.

We note that NGC~2110 was also observed as part of the Siding Spring
  Southern Seyfert Spectroscopic Snapshot
  Survey \citep[S7;][]{Thomas2017}, using 
 the WiFeS instrument on the ANU 2.3-m telescope. Although
  their FoV is larger ($25\arcsec \times 38\arcsec$) than that
  provided by MEGARA, our data have higher spatial resolution due
  to both the smaller MEGARA fiber diameter ($0.62\arcsec$) 
  versus the S7 spaxel ($1\arcsec\times 1\arcsec$) and better seeing
  conditions (MEGARA $0.9\arcsec$ versus S7 $1.4\arcsec$). In addition,
  our MEGARA observations have higher spectral resolution in
  the blue range (5000 versus 3000 for the S7 survey).

\begin{figure}[h]
  \centering
  \includegraphics{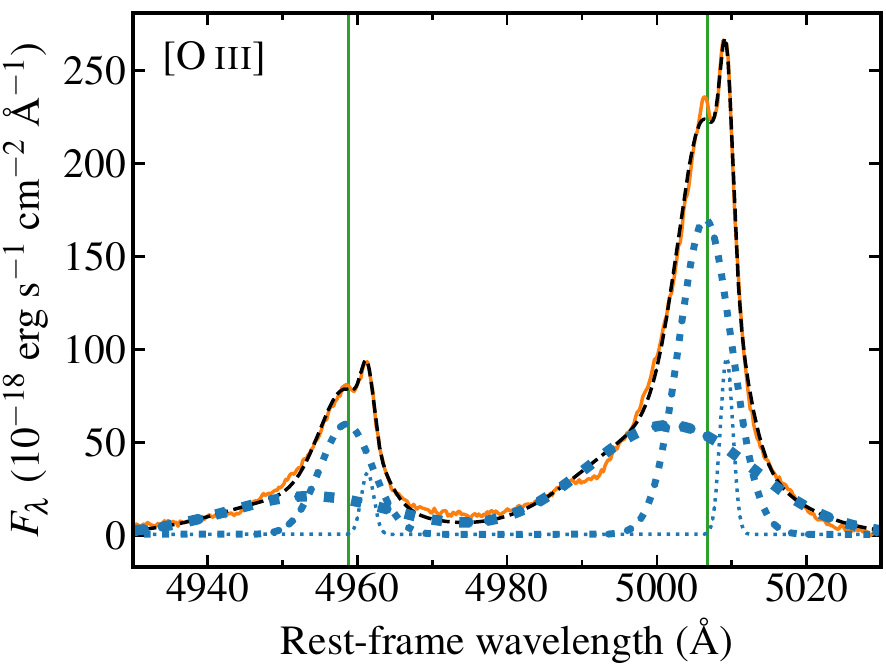}
  \caption{GTC/MEGARA nuclear spectrum of NGC~2110 in the
    spectral region around the [\ion{O}{III}]$\lambda\lambda$4959,
    5007 doublet.  The orange line is
    the observed spectrum after
      the subtraction of the continuum 
    and extracted with a $0.9\arcsec \times 0.9\arcsec$ aperture. The fit (black
    dashed line) is the sum of three Gaussian components.  The blue dotted lines
    show each component of the fit, with lines of increasing thickness for the
    components with increasing velocity dispersions. The green lines are the
    line wavelengths.}
  \label{fig:spec-center-oiii}
\end{figure}

\begin{figure*}
  \centering
  \includegraphics{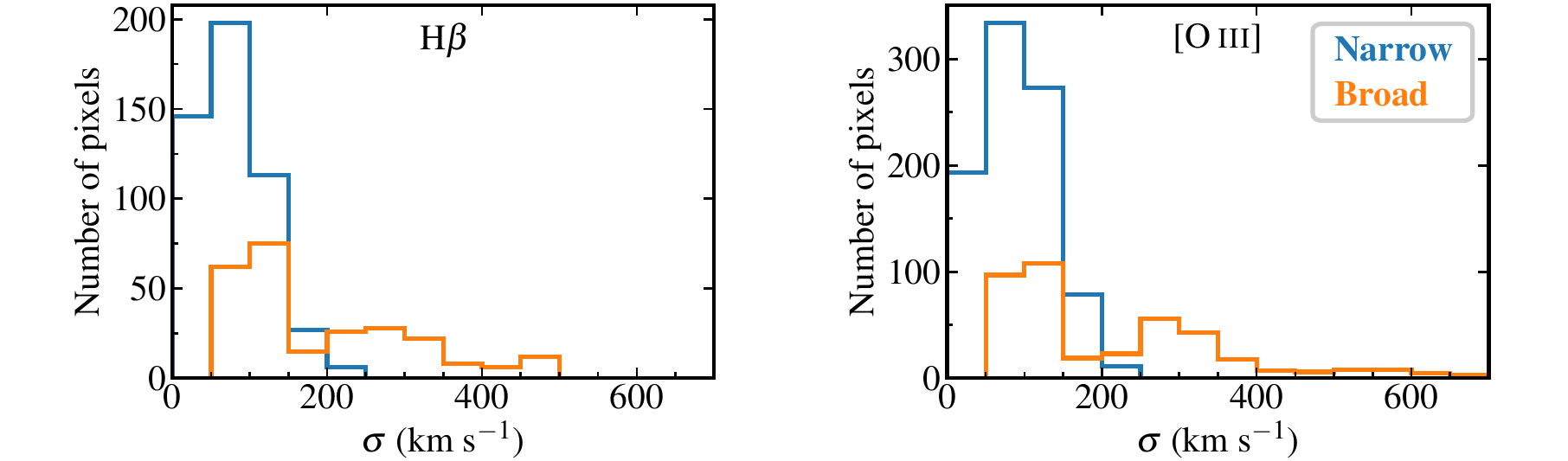}
  \caption{Observed distribution of the velocity dispersions (corrected for the
    instrumental resolution) fitted on a spaxel-by-spaxel basis and using a
    maximum of two Gaussian components from lines observed with the GTC/MEGARA
    LR-B grating, i.e., H$\beta$ and [\ion{O}{III}]$\lambda\lambda$4959, 5007
    lines. Blue and orange lines correspond to the narrow and broad components,
    respectively (see text).}
  \label{fig:sigma-hist-blue}
\end{figure*}



As the absolute flux calibration has a direct impact on the outflow properties
(Sect.~\ref{sec:outflow}),  we compare our calibration with those
provided by other authors. \citet{Davies2020} presented Very Large Telescope (VLT)
X-Shooter observations of NGC~2110 and measured
observed fluxes (not corrected for extinction) of 7.85 $\pm$ 0.10 $\times$
$10^{-15}$ erg s$^{-1}$ cm$^{-2}$ and 41.1 $\pm$ 0.5 $\times$ $10^{-15}$ erg
s$^{-1}$ cm$^{-2}$ for the H$\beta$ and H$\alpha$ lines for an aperture of
$1.8\arcsec \times 1.8\arcsec$ under seeing conditions ranging between
$0.50\arcsec$ and $0.58\arcsec$. We reproduced their aperture on our GTC/MEGARA
data and we obtained 25.6 $\times$ $10^{-15}$ erg s$^{-1}$ cm$^{-2}$ and 101.8
$\times$ $10^{-15}$ erg s$^{-1}$ cm$^{-2}$, respectively. \citet{Thomas2017}, as
part of the S7 survey, reported a dereddened flux of 384.8 $\times$
$10^{-15}$ erg s$^{-1}$ cm$^{-2}$ for the H$\beta$ line for a circular
$4\arcsec$ diameter aperture and a seeing of $1.4\arcsec$. Undoing the
dereddening in the same way as these authors (i.e., a \citealt{Fitzpatrick1999}
reddening law with $R_V$=3.1 and $A_V$=2.39\,mag) leads to a non-dereddened flux
of 29.5 $\times$ $10^{-15}$ erg s$^{-1}$ cm$^{-2}$. We extracted a spectrum with
the same aperture from our IFU data and we obtained a flux of 45.1 $\times$
$10^{-15}$ erg s$^{-1}$ cm$^{-2}$.

It is worth mentioning that both works subtracted the stellar continuum from
their spectra before measuring the line fluxes. Consequently, we also
subtracted the stellar continuum around H$\beta$ line for the above estimates
(see Sect.~\ref{sec:analysis}), while that correction was assumed negligible for
H$\alpha$. However, we noticed that the selection of stellar continuum template
contributes significantly to these differences. For instance, the above value
for the comparison with \citet{Thomas2017} was obtained using the stellar fit by
\citet{Burtscher2021}, while using the one from \citet{Davies2020} leads to a
non-dereddened H$\beta$ flux of 38.1 $\times$ $10^{-15}$ erg s$^{-1}$ cm$^{-2}$,
and when the stellar continuum is not subtracted, we found 37.7 $\times$
$10^{-15}$ erg s$^{-1}$ cm$^{-2}$ (see Sect.~\ref{sec:analysis} for more details
about stellar continuum templates). In summary, our absolute flux calibration
is 2.5 and 3.3 times higher than the estimates by \citet{Davies2020} for the
H$\beta$ and H$\alpha$ lines, respectively. However, it is \emph{only} 20-50 per
cent higher for H$\beta$ than that of the larger apertures used by
\citet{Thomas2017}. Part of the differences with \citet{Davies2020} might be
attributed to centering differences.

Additionally, we checked that our relative flux calibration within each of the
GTC/MEGARA gratings provides similar line ratios to those in
\citet{Davies2020}. We measured [\ion{O}{III}]$\lambda$5007/H$\beta$=4.96 and
[\ion{N}{II}]$\lambda$6583/H$\alpha$=1.62 using the same aperture as these
authors, while they reported a value of 4.76 and 1.37 for
these same ratios.
In what follows, we will use our own flux calibration throughout this work and
take into account the uncertainty in the absolute calibration when needed.


\subsection{Ancillary radio observations} \label{subsec:radio_data}

In order to interpret our MEGARA observations and their potential relationship
with the radio jet of NGC~2110, we downloaded from NASA/IPAC Extragalactic
Database (NED)\footnote{\url{http://ned.ipac.caltech.edu/}} a radio map observed
at 6 cm (4885 MHz) by \citet{Ulvestad1983,Ulvestad1984,Ulvestad1989}. These
observations were taken on 15$^\mathrm{th}$ March 1982, using the A
configuration of the Very Large Array (VLA) at the National Radio Astronomy
Observatory. The observations consisted of two exposures of 12 minutes
each. Calibration and self-calibration of the data followed the standard
procedures \citep[e.g.,][]{Wilson1982}. The radio map has a half-power beam
width of $0.44\arcsec \times 0.35\arcsec$ at a PA$=-1\degr$.

\citet{Ulvestad1983} found that the optical nucleus \citep[observed
  by][]{Clements1983} and the central radio component were
  coincident.  We retrieved an optical image  of NGC~2110 from ESASky
    \citep{Baines2017, Giordano2018}. The observations were taken with the Hubble
    Space Telescope ({\it HST}) WFPC2 instrument using the F791W filter. We
    adjusted the {\it HST} astrometry using the Gaia information from stars
    in the field and found that the positions of the nuclear radio and
    optical peaks agreed with each other, as can be seen from
    Fig.~\ref{fig:photo}. This fact allowed
us to assume that both the MEGARA continuum peak and the
VLA peak mark the position of the AGN.


\section{Analysis of the MEGARA observations} \label{sec:analysis}


\subsection{Emission line fitting}\label{subsec:fitting}

For the analysis of the emission lines of the MEGARA IFU data of NGC~2110, we
developed a new Python tool named \textsc{ALUCINE}\footnote{Available
  on Gitlab at  https://gitlab.com/lperalta\_ast/alucine}. This Spanish acronym 
stands for \emph{Ajuste de L\'{\i}neas para Unidades de Campo Integral de
Nebulosas en Emisi\'on} or IFU line fitting for  emission-line
  nebulae. This is a general purpose tool designed to fit emission lines in
integral-field data, using a user-specified number of components. At each
spaxel, the fit of the line (or lines) is performed automatically and
independently. We already used this tool with IFU observations obtained with the
MIRI instrument on the James Webb Space Telescope
\citep[see][]{GarciaBernete2022}.

Before we fitted the emission lines of the MEGARA observations, we subtracted
the stellar continuum around the H$\beta$ region (i.e,. within the
  wavelength range stated for that line in Table~\ref{tab:cubelets}) from the
LR-B cube on a spaxel-by-spaxel basis. This is especially important for the
H$\beta$ emission line fluxes since they are affected by stellar absorption,
while it was assumed negligible for the other emission lines. We tried three
different approaches, namely, subtracting the stellar templates provided by
\citet{Burtscher2021} and \citet{Davies2020}, and not subtracting any
template. In particular, when subtracting a stellar template we scaled
  it to match the continuum level measured in the H$\beta$ continuum region
  (defined in Table~\ref{tab:cubelets}) for each spaxel. For the figures
included in this work and all the subsequent analysis, we made the correction
with the template provided by \citet{Burtscher2021}. However, it is
  worth mentioning that we checked that results of this work are not affected
if we used the template from \citet{Davies2020}  or even if we did not
subtract the stellar continuum. For the LR-R data cube, we did not
subtract a stellar template. 
However, the  impact of this correction on the  H$\alpha$ fluxes and [N\,{\sc
  ii}]/H$\alpha$ line ratios is reduced when compared with  H$\beta$ and line ratios
involving this line because
H$\alpha$ is brighter. Moreover, \cite{Bellocchi2019} demonstrated that
the effects of the stellar template on the gas H$\alpha$ kinematics were not critical.  

In what follows, we describe the steps
followed to perform the line fitting of our MEGARA observations.

\begin{enumerate}
  \item We spatially smoothed the cubes. Specifically, we assigned to each
    spaxel the mean value in a square box of 3 spaxels ($0.9\arcsec \times
    0.9\arcsec$) around it. We chose this size for the box because it
    corresponds to the seeing value of the observations.
  \item We built cubelets from the MEGARA cubes, i.e., we created smaller cubes
    by cutting spectral slices of the cubes covering the emission lines to be
    analysed. The resulting five cubes and their wavelength ranges are listed in
    Table~\ref{tab:cubelets}.
  \item We estimated the noise in each spaxel of each cubelet using regions
    where the continuum level is flat. These continuum regions are detailed in
    Table~\ref{tab:cubelets}.
  \item Given the good quality of the GTC/MEGARA observations,
    we fitted independently the emission line(s) at each spaxel of
    each 
    cubelet. Thus, the method does not assume that all the
      emission lines share the same kinematics. We allowed for a
    maximum of two Gaussian components. 
\end{enumerate}

\begin{figure*}
  \centering 
  \includegraphics[width=17cm]{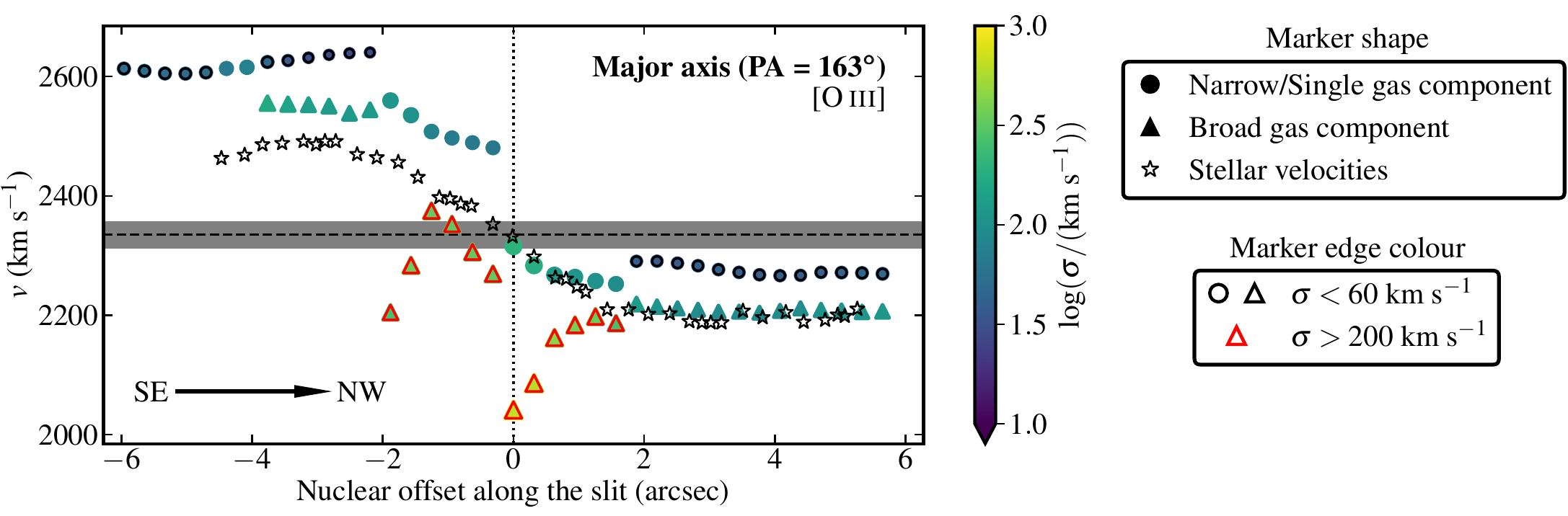}
  \includegraphics[width=17cm]{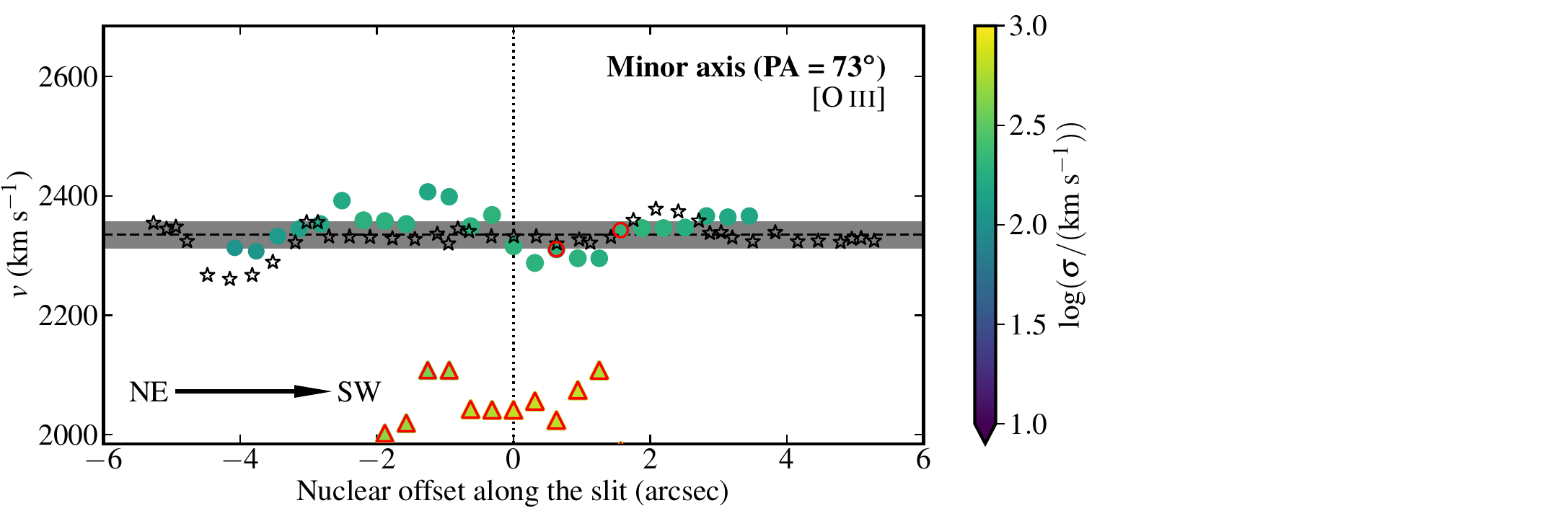}
  \includegraphics[width=17cm]{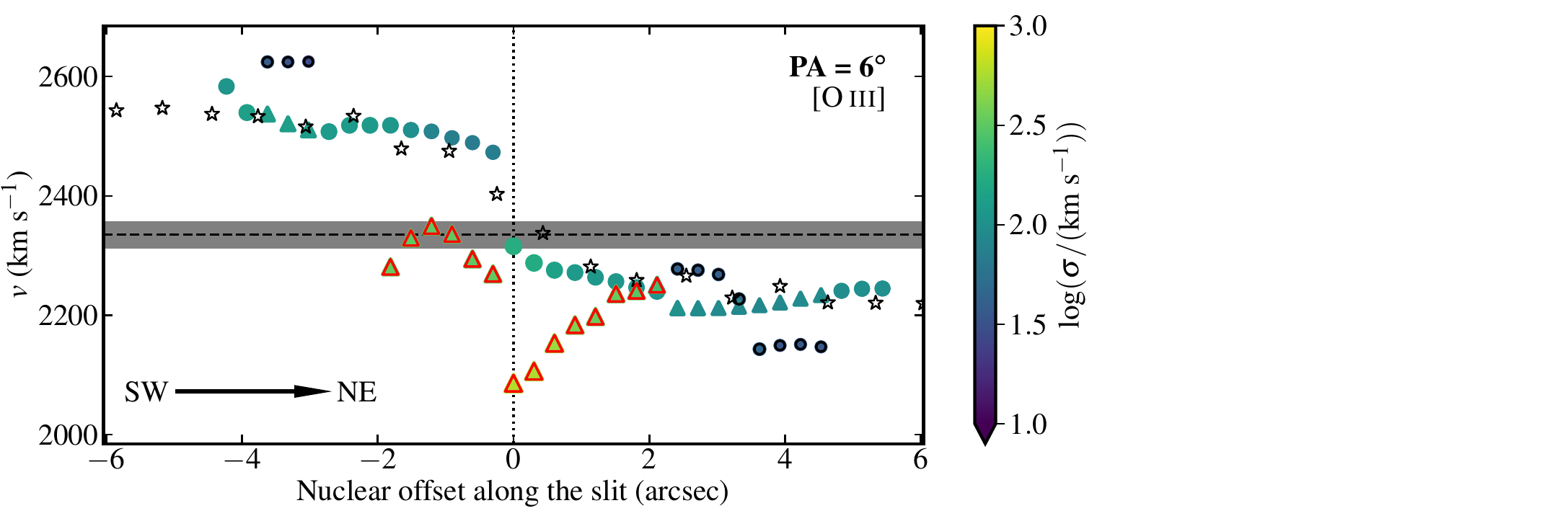}
  \caption{Comparison between the velocities of the ionised gas derived from
    fits using a maximum of two Gaussian components for the
    [\ion{O}{III}]$\lambda\lambda$4959, 5007 doublet and the stellar
    velocities. From top to bottom, the stellar velocities are along the major
    and minor axes of the galaxy from \citet{Ferruit2004} and along PA=$6^\circ$
    from \citet{GonzalezDelgado2002}. Circles represent the narrow component or
    that when only one Gaussian is required to fit the line, while triangles
    correspond to the broad component. Stars show stellar velocities. The colour
    and size of circles and triangles depend on the velocity dispersion of each
    component. Black/Red edges have been added to the circles/triangles for
    velocity dispersions below/above 60/200 km s$^{-1}$, in order to illustrate
    the component classification explained in Sect.~\ref{subsec:fitting}. The
    horizontal dashed line shows the systemic velocity, while the horizontal
    grey region its error.}
  \label{fig:velocity-slits}
\end{figure*}

In \textsc{ALUCINE}, the number of components to fit the emission lines at each
spaxel is decided as follows. We started with zero components by default. If we
detected a line above a specified threshold on amplitude over noise (AoN, see
below), we tried to fit it and we trusted the resulting fit if it had a
specified value of the minimum amplitude ($A_\mathrm{min}$). In the case of
having added one component, we iterated the criteria starting from the residual
derived from the previous fit. The values of AoN and $A_\mathrm{min}$ were tuned
for each cubelet, seeking a compromise between maximising the signal and
minimising the noise in the resulting maps. The final values of these two
parameters for each cubelet are listed in Table~\ref{tab:cubelets}. We
  should highlight that, if AoN values in this table were set to a value of
  3.0, only a few spaxels would be affected by this change. This is because the cuts
  introduced by the $A_\mathrm{min}$ values do in practice also mean that
  spaxels with low AoN are not fitted. The fitting model (e.g., the number of Gaussians
or other line profiles) depends on the cubelet and the nature of each emission
line. In particular for the MEGARA data sets, we used a baseline for adjusting
the continuum level plus the following models:

\begin{itemize}
  \item H$\beta$:
    Each component for this singlet is one Gaussian with three free parameters,
    namely, amplitude, central wavelength, and line width.
  \item {[\ion{O}{III}]}$\lambda\lambda$4959, 5007:
    Each component for this doublet is a set of two Gaussians sharing the same
    kinematics and with a fixed ratio between amplitudes of
    $A_\mathrm{[\ion{O}{III}]\ \lambda4959} = 0.350
    \ A_\mathrm{[\ion{O}{III}]\ \lambda5007}$.
   \item {[\ion{O}{I}]}$\lambda$6300:
     Although this line is a doublet with that at 6364~\AA, they are well
     separated in our GTC/MEGARA spectra. Moreover, the latter line is not close
     to any other line of interest in this work. We therefore decided to fit
     only the {[\ion{O}{I}]}$\lambda$6300 line.
   \item H$\alpha$ + [\ion{N}{II}]$\lambda\lambda$6548, 6583:
     Each component for these lines is a set of three Gaussians. Two share the
     same kinematics and have an amplitude ratio of
     $A_\mathrm{[\ion{N}{II}]\ \lambda6548} = 0.340
     \ A_\mathrm{[\ion{N}{II}]\ \lambda6583}$ in order to fit the
     [\ion{N}{II}]$\lambda\lambda$6548, 6583 doublet. The third Gaussian fits
     the H$\alpha$ line with its own three free parameters. The amplitude of the
     H$\alpha$ line is that employed to evaluate if it passes the threshold
     value $A_\mathrm{min}$ explained above.
   \item {[\ion{S}{II}]}$\lambda\lambda$6716, 6731:
     Each component for this doublet is a set of two Gaussians sharing the same
     kinematics and the ratio between amplitudes of them is a free parameter.
\end{itemize}

When fitting the lines with a particular model, we imposed a
  constraint that the fitted lines
had widths equal to or higher than the instrumental resolution of our data, that
is, 0.78~\AA \, at full width at half maximum (FWHM) for the LR-B grating and
1.0--1.1~\AA \, for the LR-R grating. As \textsc{ALUCINE} also allows the user
to apply limits to the velocities and line widths, we used this feature to a fix
spurious fits of the lines in a few spaxels. This did not affect the results
when compared to those provided with the fits without this fine tuning.

We decided to use a maximum of two Gaussian components because they provided
good quality fits to the emission lines in most of the spaxels of our MEGARA
observations. In the nuclear region of NGC~2110, we nevertheless found that a
three Gaussian component fit gave a significantly better performance. In
Fig.~\ref{fig:spec-center-oiii}, we show this fit for a spectrum extracted with
a $0.9\arcsec \times 0.9\arcsec$ aperture and the
[\ion{O}{III}]$\lambda\lambda$4959, 5007 doublet. The fits for the rest of the
emission lines can be seen in Appendix~\ref{app:nucleus}.
Table~\ref{tab:nucleus} lists the fluxes, velocities, and velocity dispersions
for each of the three components for all the emission lines.

\begin{table*}
  \caption{Measurements from the nuclear spectrum using a model with three
    Gaussian components.}
  \label{tab:nucleus}
  \centering
  \begin{tabular}{l c c c}
    \hline\hline
    Emission line &
    $F_1$ / $F_2$ / $F_3$ &
    $v_1$ / $v_2$ / $v_3$ &
    $\sigma_1$ / $\sigma_2$ / $\sigma_3$\\
    &
    $(10^{-18} \ \mathrm{erg \ s^{-1} \ cm^{-2}})$ &
    $(\mathrm{km \ s^{-1}})$ &
    $(\mathrm{km \ s^{-1}})$\\
    \hline
    H$\beta$ &
     403 /  254 /  44 &
     -47 /   -3 / 137 &
     487 /  178 /  60\\
    {[\ion{O}{III}]}$\lambda$5007 &
    1632 / 1495 / 200 &
    -317 /  -27 / 144 &
     662 /  210 /  46\\
    {[\ion{O}{I}]}$\lambda$6300 &
    1131 /  504 /  93 &
     -34 /   27 / 151 &
     443 /  157 /  49\\
    H$\alpha$ &
    1182 / 1273 / 316 &
    -281 /    7 / 142 &
     339 /  172 /  55\\
    {[\ion{N}{II}]} $\lambda$6583 &
    2855 / 1558 / 325 &
    -202 /   31 / 156 &
     446 /  161 /  50\\
    {[\ion{S}{II}]} $\lambda$6716 &
     973 /  301 / 392 &
       4 /   -8 / 150 &
     264 /   90 /  62\\
    {[\ion{S}{II}]} $\lambda$6731 &
    1012 /  256 / 292 &
       4 /   -8 / 150 &
     264 /   90 /  62\\
    \hline
  \end{tabular}
  \tablefoot{$F_i$ denotes the flux of the component $i$, $v_i$ the velocity of
    the peak of that component and $\sigma_i$ its velocity dispersion (corrected
    for instrumental resolution). The three components have been sorted in this
    table based on their velocity dispersions.}
\end{table*}


\subsection{Distributions of the fitted velocity dispersions}
\label{subsec:histogram-components}

As explained in Sect.~\ref{subsec:fitting}, we fitted the optical emission lines
with one or two components on a spaxel-by-spaxel basis. We termed them narrow
and broad (see below) when two Gaussian components were fitted, and narrow if
one Gaussian component provided a good fit. Figures~\ref{fig:sigma-hist-blue}
and \ref{fig:sigma-hist-red} show the distribution of the velocity dispersions,
corrected for instrumental resolution, for H$\beta$ and
[\ion{O}{III}]$\lambda\lambda$4959, 5007, and [\ion{O}{I}]$\lambda$6300,
H$\alpha$, [\ion{N}{II}]$\lambda\lambda$6548, 6583 and
[\ion{S}{II}]$\lambda\lambda$6716, 6731, respectively.

The velocity dispersion distributions for the narrow component are narrow and
clearly unimodal, peaking at 50--100 km s$^{-1}$. These values are
  similar to the values derived from VLT/MUSE observations for other Seyfert 2
  galaxies, like NGC~5643 \citep{GarciaBernete2021} and NGC~7130
  \citep{Comeron2021}. Regarding the broad component, most of the distributions
tend to be bimodal, with one peak with velocity dispersions at $100-150$\, km
s$^{-1}$, thus broader than the narrow component. The second peak of the broad
component is at $\sigma \sim$200--350 km s$^{-1}$, followed by a tail towards
high velocity dispersions. The latter reaches values of up to $\sim$500--700 km
s$^{-1}$, mostly in [\ion{O}{III}]$\lambda\lambda$4959, 5007 and
[\ion{N}{II}]$\lambda\lambda$6548, 6583. The distributions of the fitted
velocity dispersions in the MEGARA LR-R data cube are similar to those of
\cite{SchnorrMueller2014}.


\subsection{Comparison with the stellar kinematics}
\label{subsec:compstellarkins}

To classify the different ionised gas components according to their kinematics,
in this section we compare the gas velocities with those measured from stellar
absorption features. Figure~\ref{fig:velocity-slits} shows the velocities of the
two components fitted for the [\ion{O}{III}]$\lambda$5007 line versus the
stellar kinematics along three different slit PAs. The top and middle panels are
the stellar kinematics along the galaxy major (PA=$163^\circ$) and minor
(PA=$73^\circ$) axes from \cite{Ferruit2004}, while the bottom panel is along
PA=$6^\circ$ from \cite{GonzalezDelgado2002}. For reference, we
  show the orientation of these slits in Fig.~\ref{fig:photo}.

 Figure~\ref{fig:velocity-slits}   already reveals the
difficulty in classifying the different ionised gas kinematic components in
NGC~2110. Along the NGC~2110 major axis, our MEGARA observations detected
two components out 
to a radial distance of approximately $4\arcsec$ to the southeast and $6\arcsec$
to the northwest. Along the minor axis, the two components are seen in the inner
$\sim4\arcsec$. In this central region and along the minor axis, the
[\ion{O}{III}] narrow component follows relatively well the stellar kinematics
(middle panel of Fig.~\ref{fig:velocity-slits}). When comparing with the
stellar
kinematics of \citet{GonzalezDelgado2002} along PA=$6^\circ$, the ionised
gas component with the intermediate velocity dispersions agrees well with the
stars (bottom panel of Fig.~\ref{fig:velocity-slits}).

Along the major axis of the
galaxy  to  the northwest and in the central $\sim 2\arcsec$, the
  velocities of the gas narrow component are comparable to those of
  the stars,  while beyond $2\arcsec$ they are redshifted. To the
    southeast the gas velocities are redshifted with respect to those
    of the stars (top
panel of Fig.~\ref{fig:velocity-slits}). This is similar to the findings of
\citet{Ferruit2004}. The broad component with $\sigma \simeq
250-670\,\mathrm{km\,s}^{-1}$ in the inner $4\arcsec$ is likely tracing an
outflow, which we will analyse in more detail in
Sect.~\ref{sec:outflowcomponent}. In the outer regions to the northwest and
along the galaxy major axis, the component with intermediate values of velocity
dispersion traces well the stellar kinematics.
The narrowest gas component (circles in the figure)  to the southeast at
radial distances $>2\arcsec$ deviates the most from the stellar
velocities, with gas velocities faster than the stellar ones.

A plausible scenario accounting for a slower rotation of the stars
  relative to the gas in some parts of the galaxy would invoke a rotation lag in thick stellar
  discs or a comparatively larger contribution of the velocity
  dispersion in the stellar component. However, the behaviour is the
  opposite in the northwest side. Here, the reported asymmetry
  of the gas kinematics in NGC~2110 could be explained by an uneven
  impact left by the radio jet on the gas motions,
  which probably
  slowed down the gas with the lowest observed velocity dispersions
  on the northern side of the disc. The secondary peak of the ALMA
  millimeter emission \citep[see panel (c) of Figure~1
  of][]{Rosario2019} identifies the region where the jet was probably diverted.
  We will discuss the impact of the radio jet in more detail 
  in Sect.~\ref{subsec:outflow-jet}.

\begin{figure*}
  \centering
  \includegraphics[width=17cm]{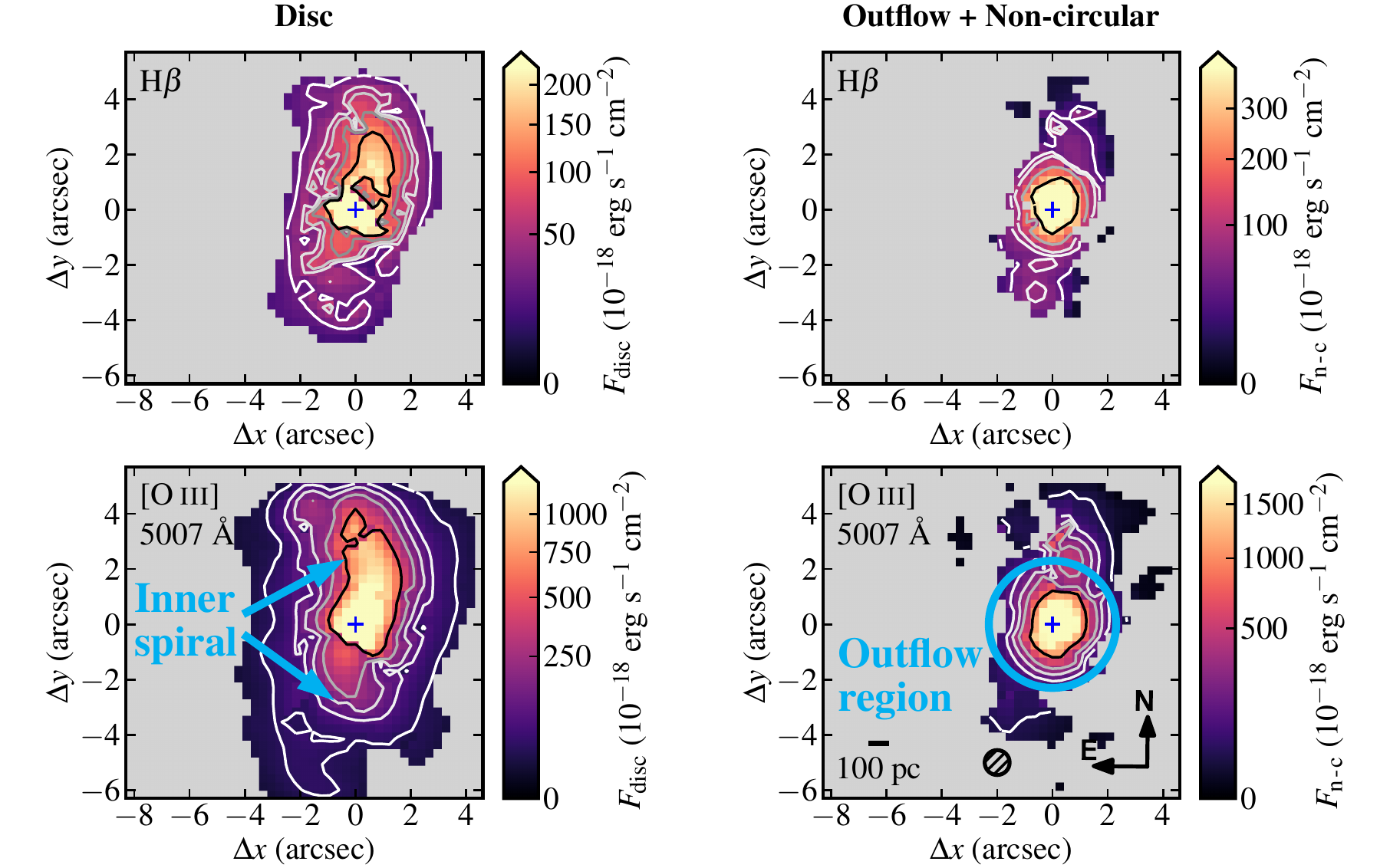}
  \caption{Flux maps of the disc (left) and the outflow+non-circular (right)
    components for the lines observed with the GTC/MEGARA LR-B grating, i.e.,
    the H$\beta$ and [\ion{O}{III}]$\lambda$5007 lines. All the colour bars use
    a square root scale and are saturated at the half maximum value of each map.
    Contour levels correspond to 30, 60, 70, 80 and 90 per cent of the maximum
    value of each map. Crosses indicate the location of the nucleus (as defined
    by the peak of the continuum). The north-east compass, the hatched
      circle indicating the seeing size, and physical size bar at the bottom
    right map apply to all the maps.}
  \label{fig:flux-blue}
\end{figure*}

\begin{figure*}
  \centering
  \includegraphics{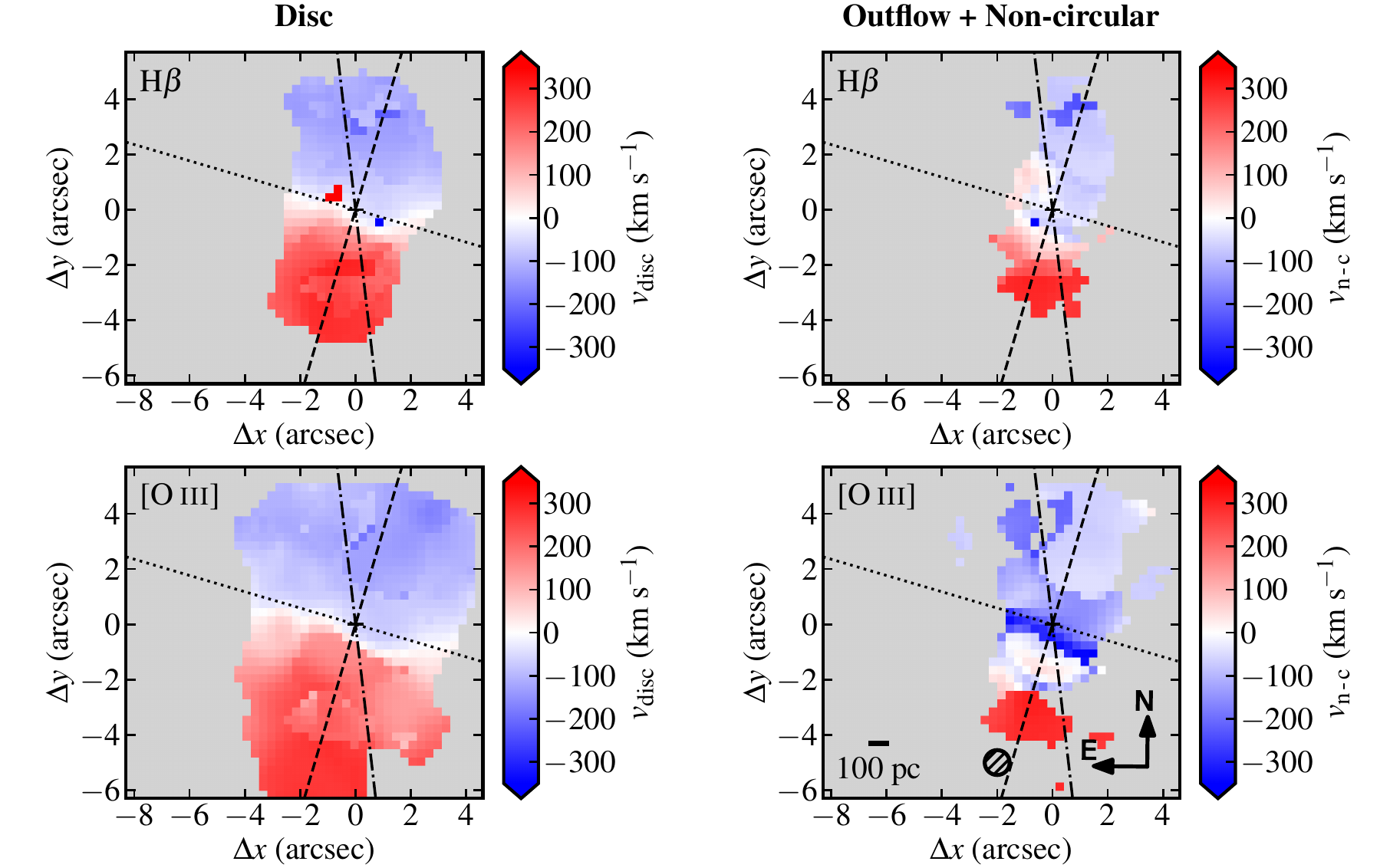}
  \caption{Velocity maps of the disc (left) and the outflow+non-circular (right)
    components for the lines observed with the GTC/MEGARA LR-B grating, i.e.,
    H$\beta$ and [\ion{O}{III}]$\lambda\lambda$4959, 5007 lines. Values are
    referred to the systemic velocity. The lines indicate the orientation of the
    slit PA in Fig.~\ref{fig:velocity-slits} (see text for details). Other
    symbols as in Fig.~\ref{fig:flux-blue}.}
  \label{fig:velocity-blue}
\end{figure*}

Based on this comparison, where we identify the rotating gas with the intermediate
  velocity dispersion component, we define the following components for the ionised
gas:

\begin{itemize}
  \item \emph{Disc component}:
    At a particular spaxel, if only one component was required to fit the
    emission line, this is identified with the disc. If there are two
    components, we apply the following criteria for assigning the component
    whose kinematics is closer to the stellar one: (i) If both are below $\sigma
    = 200\,\mathrm{km\,s}^{-1}$ (for the [\ion{O}{III}] doublet), the broader
    component is assigned to the disc, (ii) If one of them is above
    $200\,\mathrm{km\,s}^{-1}$ (for the [\ion{O}{III}] doublet), the narrow
    component is assigned to the disc. It is worth mentioning that
    we detected no spaxels in which both components were broader than
  $200\,{\rm km\,s}^{-1}$, nor in any of the one-component
      spaxels did the velocity dispersion
      exceed $200\,{\rm km\,s}^{-1}$. The typical values for the velocity
    dispersion of the disc component are $60-200\,\mathrm{km\,s}^{-1}$.
  \item \emph{Outflow component}:
    This component includes those spaxels where one of the components of the
    [\ion{O}{III}] doublet was fitted with a velocity dispersion above $\simeq
   200\,{\rm km\,s}^{-1}$. Most of these spaxels lie in the central
    4\arcsec.
  \item \emph{Non-circular component}:
    If at a particular spaxel the line requires two components, the
      component which is not assigned to the disc is termed
      non-circular. We found that these components lie outside
        the outflow region and generally include spaxels whose lines
        were fitted with low values of the velocity dispersion 
    (<60 km s$^{-1}$).
\end{itemize}

\begin{figure*}
  \centering
  \includegraphics{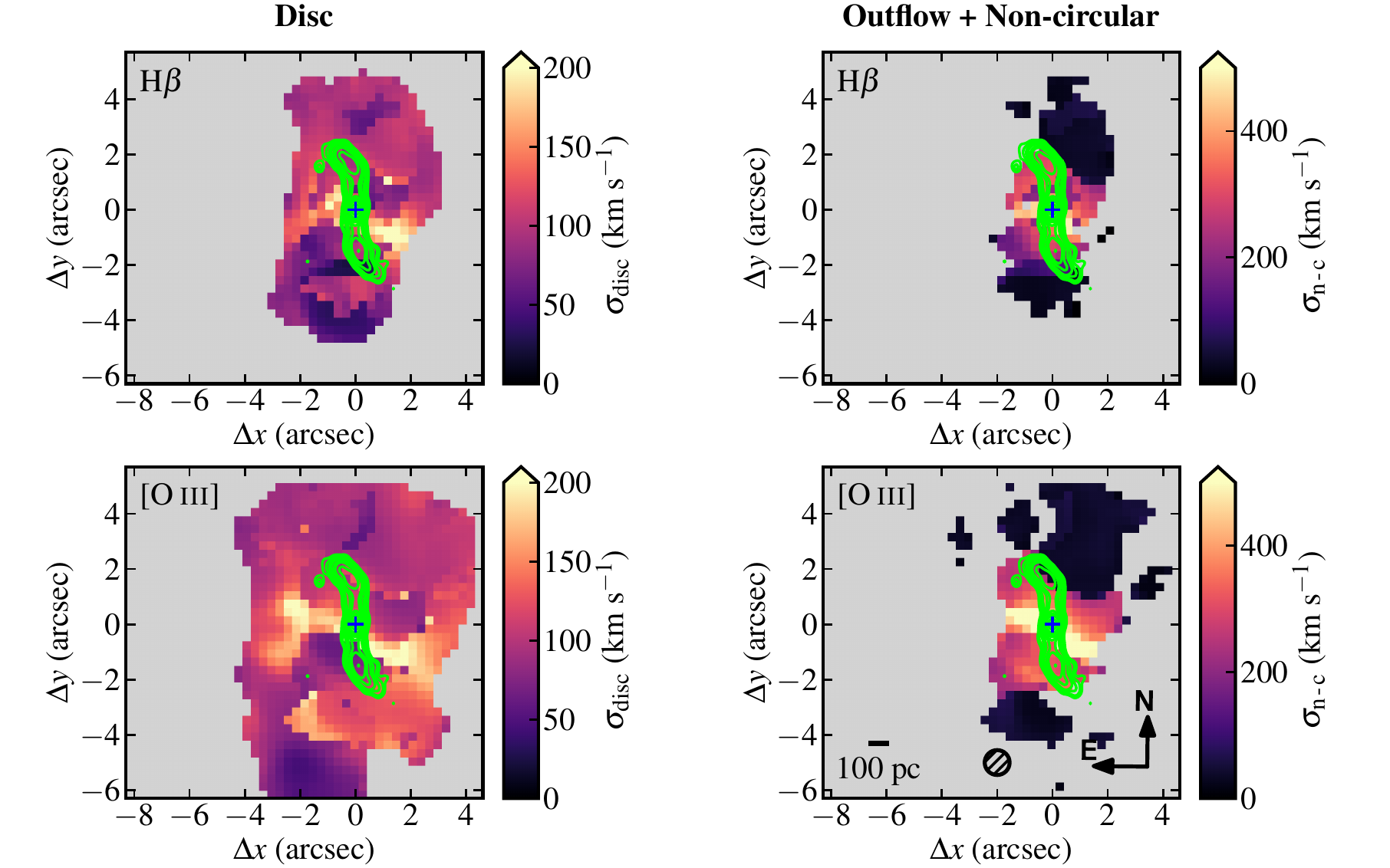}
  \caption{Velocity dispersion maps of the disc (left) and the
    outflow+non-circular (right) components for the lines observed with the
    GTC/MEGARA LR-B grating, i.e., H$\beta$ and
    [\ion{O}{III}]$\lambda\lambda$4959, 5007 lines. Contours show the VLA radio
    jet (Sect.~\ref{subsec:radio_data}). Symbols as in
    Fig.~\ref{fig:flux-blue}.}
  \label{fig:sigma-blue}
\end{figure*}

This component classification differs from that proposed by
\citet{SchnorrMueller2014} for NGC~2110. Although for each spaxel they also
fitted the emission lines in the red part of the spectrum with two components,
they classified them into four categories only according to increasing velocity
dispersion values, namely, outflow, warm gas disc, cold gas disc, and a north
cloud. With this approach, they sought to interpret the asymmetry in the
velocity field and proposed two \emph{incomplete} rotating disc components (see
their Figs.~5 and 6). Our kinematic classification  has only \emph{one} disc
component, which includes a large fraction of the spaxels in their cold
and warm disc components. On the other hand, we agreed with their velocity
dispersion classification for the outflow component. We nevertheless found that,
for the goals pursued by this work, it is sufficient to perform a classification
which allows us to separate the outflow component from that of the disc and
other non-circular components.


\subsection{Two-dimensional maps}\label{subsec:maps}

With the kinematics classification adopted in the previous section, we created
maps of the line intensity, velocity, and velocity dispersion for the disc
component and the outflow+non-circular component. For the latter, the outflow
component is confined to the central $4\arcsec$ and thus is easily distinguished
from the other non-circular motions by inspecting the velocity dispersion
map. In Figs.~\ref{fig:flux-blue}, \ref{fig:velocity-blue}, and
\ref{fig:sigma-blue}, we show the MEGARA LR-B grating maps for the intensity,
velocity, and velocity dispersion, respectively for H$\beta$ and
[\ion{O}{III}]. These maps capture the most remarkable aspects of both
gratings. The LR-R grating maps can be found in Appendix~\ref{app:lr-r}.

Diagnostic diagrams (also known as BPT diagnostic diagrams) based on optical
line ratios are tools to identify the dominant gas excitation mechanism in
galaxies. Initially proposed by \citet{Baldwin1981} and \citet{Veilleux1987},
they use a comparison of the [\ion{O}{III}]/H$\beta$ ratio versus the
[\ion{S}{II}]/H$\alpha$, [\ion{O}{I}]/H$\alpha$ and [\ion{N}{II}]/H$\alpha$
ratios. These diagrams can be used to distinguish between mechanisms such as
photoionisation by young stars in star forming regions, photoionisation
from an AGN, or low ionisation (nuclear) emission regions (LI(N)ER). The
  last type  is associated with the presence of shocks, as well as
  ionisation by hot, low-mass evolved stars (see also
  Sect.~\ref{sec:disccomponent}).
BPT diagrams derived from IFU
spectroscopy are routinely applied to local AGN
\citep[e.g.,][]{Bremer2013,Davies2016,Husemann2019,Mingozzi2019,DAgostino2019,
  SmirnovaPinchukova2019,Shimizu2019, Perna2020, Venturi2021,Comeron2021,
  SmirnovaPinchukova2022,Juneau2022,Cazzoli2022}. Figures~\ref{fig:bpt-disc} and
\ref{fig:bpt-out} display the [\ion{O}{III}]/H$\beta$ versus
[\ion{N}{II}]/H$\alpha$, [\ion{S}{II}]/H$\alpha$, and [\ion{O}{I}]/H$\alpha$
diagnostic diagrams, and their associated spatially-resolved maps, for the disc
and outflow+non-circular components, respectively.

Finally, we computed electron density maps using the fluxes from the
[\ion{S}{II}] $\lambda\lambda$6716, 6731 doublet. In particular, we used the
method \texttt{Atom.getTemDen} of \textsc{PyNeb} \citep{Luridiana2015}, and
assumed a typical electron temperature of $10^4$ K. This method computes
densities in the range 20 -- 5.7 $\times$ $10^4$ cm$^{-3}$ for
[\ion{S}{II}]$\lambda$6731/[\ion{S}{II}]$\lambda$6716 ratios between 0.7 and
2.2. We assigned the minimum/maximum density to those ratios which are
below/above the mentioned interval. However, the upper value is only reached at
one spaxel within the central region and at a spurious location of the
non-circular component. The resulting electron density maps are in
Fig.~\ref{fig:density}. For comparison, we also estimated the electron densities
using the ionisation parameter (U) method proposed by \citet{Baron2019}. This
method only applies to gas excited by a central source (like an AGN). Given the
likely mix of shock and AGN ionisation in some regions of NGC~2110 (see below),
the interpretation of the electron density maps using the U method are not
straightforward. The maps are shown in the Appendix in
Fig.~\ref{fig:density-log_u}.


\section{The disc component} \label{sec:disccomponent}

The maps of the emission line fluxes for the disc component are the left panels
of Figs.~\ref{fig:flux-blue} and \ref{fig:flux-red-a} for H$\beta$ and
[\ion{O}{III}]$\lambda$5007, and [\ion{O}{I}]$\lambda$6300, H$\alpha$,
[\ion{N}{II}]$\lambda$6583, [\ion{S}{II}]$\lambda$6716 and
[\ion{S}{II}]$\lambda$6731, respectively. We detected the disc component filling
approximately the full MEGARA FoV along the major axis of the galaxy and
approximately $8\arcsec$ along the minor axis of the galaxy. The ionised gas
morphology of the disc component is dominated by the nucleus, with the extended
emission tracing the disc orientation and some of the spiral morphology (marked
in the bottom left panel of Fig.~\ref{fig:flux-blue}), seen in the
\emph{HST} image in Fig.~\ref{fig:photo} and
especially in the optical color maps \citep[see for instance, Fig.~1
  of][]{SchnorrMueller2014}.

Figures~\ref{fig:velocity-blue} and \ref{fig:velocity-red} (left panels) are the
velocity maps of the disc component. They all show a similar structure, namely a
rotation pattern in which the north half approaches the observer and the south
half recedes from the observer. The absolute values of velocities in the north
half are lower than those in the south half. Our results for this component are
in line with the observations by \citet{Wilson1985},
\citet{GonzalezDelgado2002}, \citet{Ferruit2004}, and \citet{Rosario2010}. A few
spaxels close to the nucleus have high negative velocities in the maps from the
[\ion{N}{II}]$\lambda\lambda$6548, 6583 and [\ion{S}{II}]$\lambda\lambda$6716,
6731 doublets, and correspond to spurious fits of our automatic algorithm. We
also observe some discontinuities in the velocity and velocity dispersion maps,
especially in transition regions between lines fitted with two components and
only one component. It is possible that in some spaxels, in particular in the
central and transition regions, the lines present other additional velocity
components that cannot be resolved with the current spectral resolution.

\begin{figure*}
  \centering
  \includegraphics{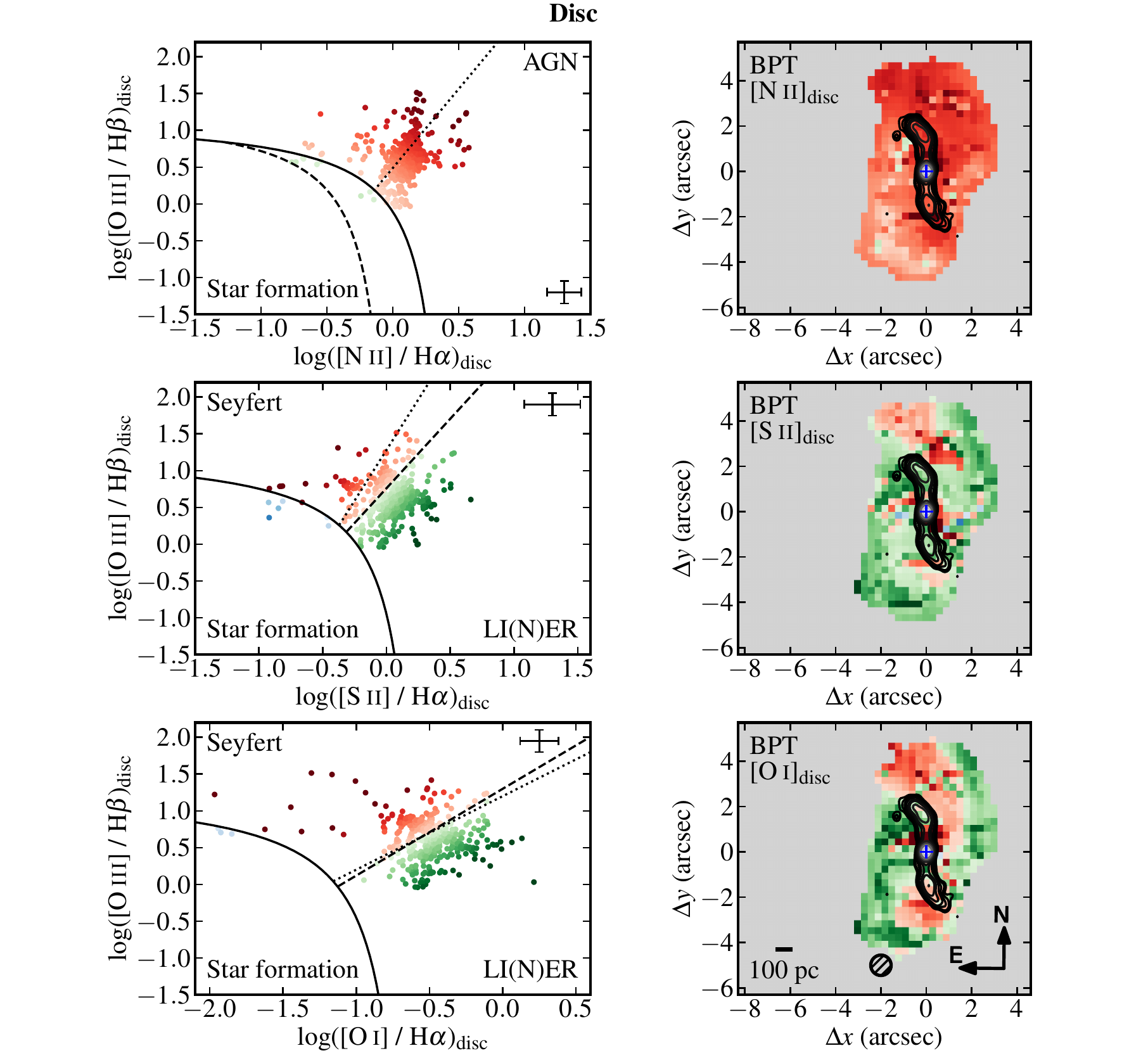}
  \caption{Spatially-resolved BPT diagnostic diagrams (left panels) and maps
    (right panels) for the disc component. Top, middle and bottom diagrams of
    the left panels correspond to [\ion{O}{III}]$\lambda$5007/H$\beta$ versus
   [\ion{N}{II}]$\lambda$6583/H$\alpha$, ([\ion{S}{II}]$\lambda$6716 +
    $\lambda$6731)/H$\alpha$ and [\ion{O}{I}]$\lambda$6300/H$\alpha$,
    respectively. The points in the red color scale are spaxels in the
    AGN/Seyfert region, those in the blue color scale are spaxels in the
      star-forming region, and those in the green color scale in the
      composite-like/LI(N)ER region. Solid curves are theoretical upper limits
    of starburst models from \citet{Kewley2001}. The dashed curve in the
    [\ion{N}{II}] diagram is the demarcation between starburst galaxies and AGN
    proposed by \citet{Kauffmann2003}. Dashed lines in the [\ion{S}{II}] and
    [\ion{O}{I}] diagrams are classification lines proposed by
    \citet{Kewley2006} to split Seyfert and LI(N)ER emission. Dotted lines in
    the diagnostic diagrams mark bisector lines between the loci of points for
    the shock-excited NGC~1482 and the AGN-excited NGC~1365
    \citep{Sharp2010}. For reference, the error bars represent the measurement
      uncertainties in a
      spaxel outside the outflow region, at $\simeq 2.4\arcsec$ north of the AGN. The right panels
    are diagnostic maps colour coded as 
    their corresponding diagrams to the left. Contours in maps show the radio
    jet using the VLA data described in Sect.~\ref{subsec:radio_data}.}
  \label{fig:bpt-disc}
\end{figure*}

\begin{figure*}
  \centering
  \includegraphics{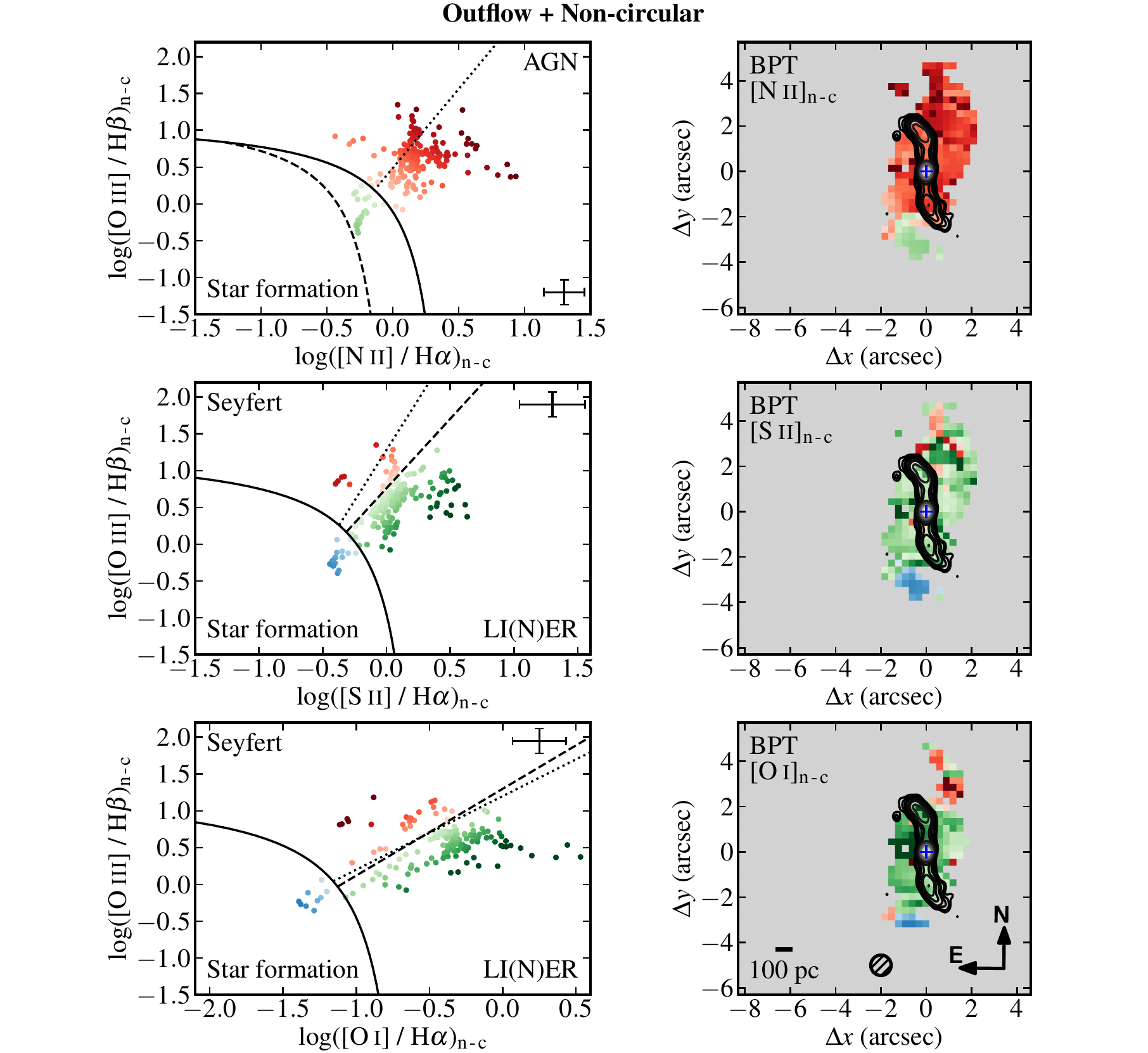}
  \caption{Spatially-resolved BPT diagnostic diagrams (left) and maps (right)
    for the outflow+non-circular component. All symbols, colors, and lines are
    as in Fig.~\ref{fig:bpt-disc}.}
  \label{fig:bpt-out}
\end{figure*}

Figures~\ref{fig:bpt-disc} and \ref{fig:bpt-out} present, for the first time,
the spatially-resolved BPT diagrams of the central part of the disc and outflow
region of NGC~2110. All the spaxels of the disc component fall in the AGN zone
in the [\ion{N}{II}] diagnostic diagram (Fig.~\ref{fig:bpt-disc}, top left
panel). There appears to be a gradient of increasing excitation
  to the north of the AGN location, which might also be caused by an uneven
  impact of the radio jet, as discussed in
  Sect.~\ref{subsec:compstellarkins}.
LI(N)ER emission predominates in the diagnostic diagrams from
[\ion{S}{II}] and [\ion{O}{I}] (Fig.~\ref{fig:bpt-disc}, middle and bottom left
panels, respectively). However, in these two last maps, there are some regions
with Seyfert-like excitation, that is, gas photoionised by the AGN. These are
near the AGN position as well as beyond the bending of the radio jet, both to
the north and south of the AGN, at projected distances reaching at least
$5\arcsec \simeq 800\,$pc. These regions appear to be coincident with the inner
spiral arms marked in Fig.~\ref{fig:flux-blue} (bottom right panel). They are
likely tracing gas in the disc that is being illuminated by the AGN, as has been
found in other Seyfert galaxies in the MAGNUM sample \citep{Mingozzi2019}. In
fact, this morphology could be understood in the context of the geometry
proposed by \cite{Rosario2010} in their Fig.~7. The Seyfert-like excitation
regions could plausibly arise by the intersection of a wider-angle ionisation
cone with the galaxy disc. This could also explain the relatively low ionisation
of the entire observable ionisation cone, as we are only seeing the part that is
on the edge of the cone due to the geometry of illumination. This region is
likely to correspond to the continuation of the AGN-photoionised plume
identified by \cite{Rosario2010}.

Since NGC~2110 is an early-type spiral galaxy, LI(N)ER-like line ratios
  may not necessarily be interpreted as evidence for shocks. To rule
  out a dominant contribution from ionisation by evolved stars, we
  constructed  diagrams comparing the equivalent width of H$\alpha$
  (W$_{{\rm H}\alpha}$)
  with the  [\ion{N}{II}]/H$\alpha$ line ratio, also known as
  WHAN diagrams \citep{CidFernandes2010,
    CidFernandes2011} for both the disc and outflow components. As can
  be seen from Fig.~\ref{fig:whan}, the majority of spaxels of the
  disc and outflow component lie outside the
  region occupied by retired and passive galaxies. According to these
  authors, the gas excitation  galaxies with very
  weak line emission is believed to be due to
  evolved stars. We can rule out that in the central disc
  and outflow regions
  of NGC~2110, the LI(N)ER-like emission is dominated by excitation
  from evolved stars. Therefore, considering the results from the
    WHAN diagrams and the seeing conditions, we can conclude that
    there is  LI(N)ER-like excitation in galaxy regions well beyond
    what could be expected from beam smearing.

\section{The outflow component} \label{sec:outflowcomponent}


\subsection{Morphology and excitation conditions}
\label{subsec:outflowmorphology}

The intensity of the outflow component of the ionised gas peaks at the AGN
position (central region in the right panels of Figs.~\ref{fig:flux-blue} and
~\ref{fig:flux-red-a}), and can be easily distinguished from the other
non-circular motions by the velocity and velocity dispersion maps (right panels
of Figs. \ref{fig:velocity-blue}, ~\ref{fig:velocity-red}, \ref{fig:sigma-blue},
and ~\ref{fig:sigma-red}). Nevertheless, we marked the outflow region in the
bottom right panel of Fig.~\ref{fig:flux-blue}. The outflow emission is oriented
along the northwest-southeast direction at a position angle that is similar to
that of $-35\degr$ observed in the X-ray Fe-K$\alpha$ line by
\cite{Kawamuro2020}. The outflow region is clearly resolved along the
north-south direction, with a projected size (FWHM) of $1.34\arcsec \simeq
214\,$pc at the [\ion{O}{III}] wavelength compared with the measured FWHM of
$0.93\arcsec$ from the calibration star at the same wavelength and direction. We
obtained these measurements from the data cubes before we spatially smoothed
them (see Sect.~\ref{subsec:fitting}). Assuming a simple broadening in
quadrature, the intrinsic size would be approximately $1\arcsec \simeq 160\,$pc
(FWHM). It is just resolved in the east-west direction with a FWHM$=1.2\arcsec$,
compared with the standard star size of FWHM$=0.96\arcsec$ along this direction.

The VLA radio emission \citep{Ulvestad1983} of NGC~2110 shows a collimated
linear structure in an almost north-south direction in the inner
$2\arcsec$. This is in agreement with the size of the outflow region. Beyond
this, the radio jet bends to the northeast and southwest direction in an
inverted `S-shaped' morphology, which coincides with the noticeable drop
observed in the MEGARA velocity dispersion maps. Within the central $\sim
2\arcsec$, the broad component flux maps \citep[see also][]{Ferruit2004} resolve
the `outer plume' (see the middle right panel of Fig.~\ref{fig:flux-red-a})
detected by \citet{Rosario2010} in the \emph{HST} narrow-band
H$\beta$+[\ion{O}{III}] and H$\alpha$+[\ion{N}{II}] images, as well as some
fainter spiral morphology seen on larger scales. The `inner plume' identified by
\cite{Rosario2010}, extends only approximately $0.4\arcsec$ from the nucleus,
and is not resolved in our MEGARA observations.

The outflow component shows AGN excitation in the [\ion{N}{II}] diagnostic
diagram (top right panel of Fig.~\ref{fig:bpt-out}) and a LI(N)ER excitation in
the other two diagrams (middle and bottom right panels of
Fig.~\ref{fig:bpt-out}), as found in other AGN \citep[see e.g.,][]{Perna2017}.
The outflow component of NGC~2110 is relatively compact and does not reveal a
clear conical shape illuminated by the central AGN. This contrasts with the
results for the Seyfert galaxies in the MAGNUM survey
\citep{Cresci2015,Mingozzi2019} and other works \citep[see e.g.,][]{Juneau2022},
in which ionisation cones with AGN photoionised emission, with their edges
showing lower ionisation, are frequently revealed in their diagnostic diagrams.

As shown above, a large fraction of the spaxels in the disc and outflow
component of NGC~2110 have line ratios in the LI(N)ER region in two BPT
diagrams, where the gas excitation can be explained by shocks \citep[see][and
  references therein]{Perna2017,Mingozzi2019,Cazzoli2022} but also by X-ray
radiation from the AGN and evolved stars. However, 
  Fig.~\ref{fig:whan} (right panels) shows that most spaxels are outside
  the region occupied by retired and passive galaxies where the LI(N)ER excitation is
  produced by evolved stars. Comparing with Fig.~8 of
\cite{Mingozzi2019}, it would be possible to reproduce the observed line ratios
for the outflow component and some regions in the disc with shock models. Such
shocks might be driven by the radio jet impacting the ISM in the galaxy
disc. Indeed, recent simulations show that low-power radio jets, such as those
observed in Seyfert galaxies, can drive $\sim$ kpc-scale outflows, which can
affect the inner galaxy disc \citep{Mukherjee2018,Talbot2022,Meenakshi2022}. The
effects are especially noticeable when jets are launched at low inclinations
with respect to the disc of the galaxy. Finally, we have only been able to
measure all the line ratios of the non-circular component in  a small number of
spaxels. The north part appears to be located in the AGN area of the
[\ion{N}{II}] diagram and the LI(N)ER region in the [\ion{S}{II}] and
[\ion{O}{I}] diagrams. The south part, at an approximate radial distance from
the AGN of $\sim 4\arcsec$, appears in the composite regime and the star forming
area in the BPT diagrams.


\subsection{Kinematics} \label{subsec:outflowkinematics}

The velocity fields of the outflow component (central region in the right panels
of Figs.~\ref{fig:velocity-blue} and \ref{fig:velocity-red}) present moderate
values of the peak velocity relative to the systemic value, typically in the
$\sim 50-100\,$km s$^{-1}$ range. The typical values of the velocity dispersion
in the outflow region are 700\,km s$^{-1}$. The
[\ion{O}{III}]$\lambda\lambda$4959, 5007, H$\alpha$ and
[\ion{N}{II}]$\lambda\lambda$6548, 6583 maps show a linear structure almost
$4\arcsec$ in size, with high blue shifts from north-east to south-west through
the nucleus reaching velocities of $\sim$300 km s$^{-1}$. This region is
coincident with a high-$\sigma$ strip seen in some of the outflow component
velocity dispersion maps (Figs.~\ref{fig:sigma-blue} and \ref{fig:sigma-red})
and is also seen in the disc component. It has already been reported by
\citet{GonzalezDelgado2002} and \citet{SchnorrMueller2014}. This high-$\sigma$
strip is close to the minor axis of the galaxy, so it is possible that smearing
of the rotation curve and other non-circular velocities could partly explain it.

The MEGARA [\ion{O}{III}]$\lambda$5007 position-velocity (p-v) diagrams along
the major and minor axis of the galaxy (Fig.~\ref{fig:pv}) are useful to
identify the velocities of the outflow. We constructed these diagrams from the
LR-B data cube after continuum subtraction and using an aperture of
$0.9\arcsec$. The diagram for the major axis can be compared with that presented
by \citet{Rosario2010} for their \emph{HST} slit-A spectroscopy (c.f. their
Fig.~3). We note that the PA of their slit is not exactly along the major axis
of the galaxy, but it is close enough for this comparison. As can be seen from
left panel of Fig.~\ref{fig:pv}, the MEGARA p-v diagram reproduces well the
\emph{HST} observations considering our $\sim0.9\arcsec$ seeing conditions. This
p-v diagram also reveals the asymmetric rotation curve discussed in
Sect.~\ref{subsec:compstellarkins}. To the southeast of the AGN, at distances of
$\simeq 2.5\arcsec$, it clearly shows the two velocity components fitted in
Fig.~\ref{fig:velocity-slits}. The outflow can be seen as much higher velocities
in the central $\simeq 2\arcsec$, both blueshifted to velocities in excess of
$-1000\,\mathrm{km\,s}^{-1}$ and redshifted to velocities of several hundreds of
km s$^{-1}$. The largest values of the blueshifted velocities are mostly
concentrated in the nuclear region while some more moderate redshifted and
blueshifted velocities are observed along the major axis to the northwest of the
AGN.

\begin{figure*}
  \centering
  \includegraphics{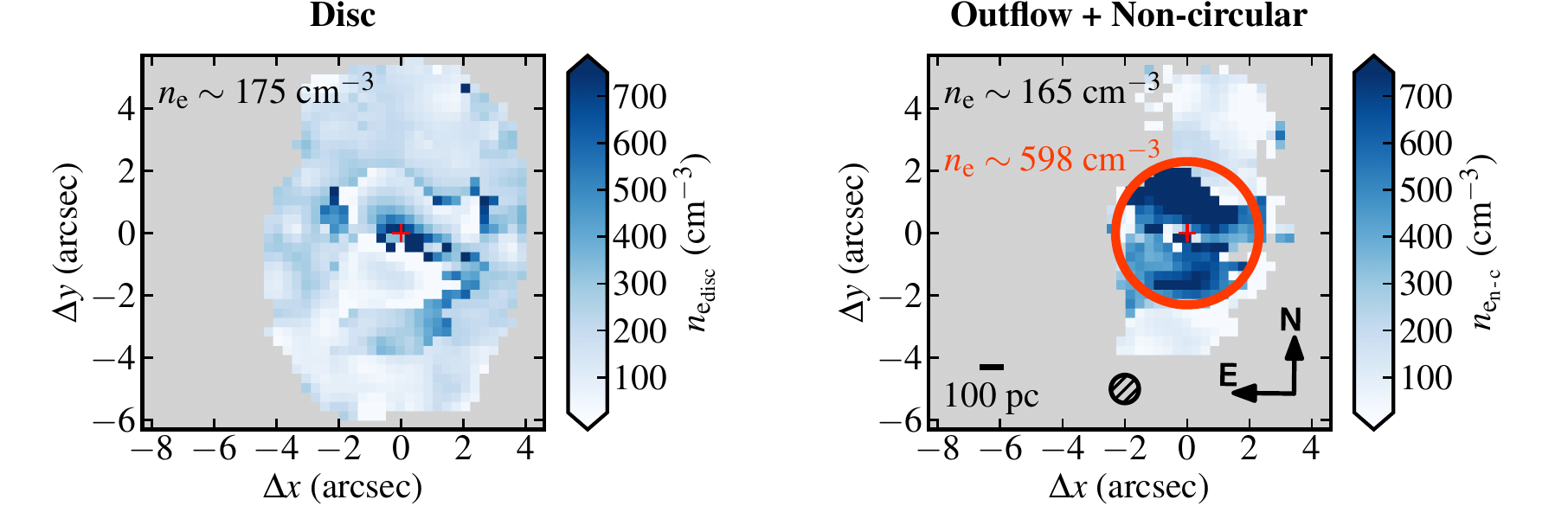}
  \caption{Maps of electron densities for the disc (left) and the
    outflow+non-circular (right) components using the [\ion{S}{II}] doublet
    ratio method. Values on top left corners indicate the median density in each
    map.}
  \label{fig:density}
\end{figure*}

\begin{figure*} 
  \includegraphics[width=18cm]{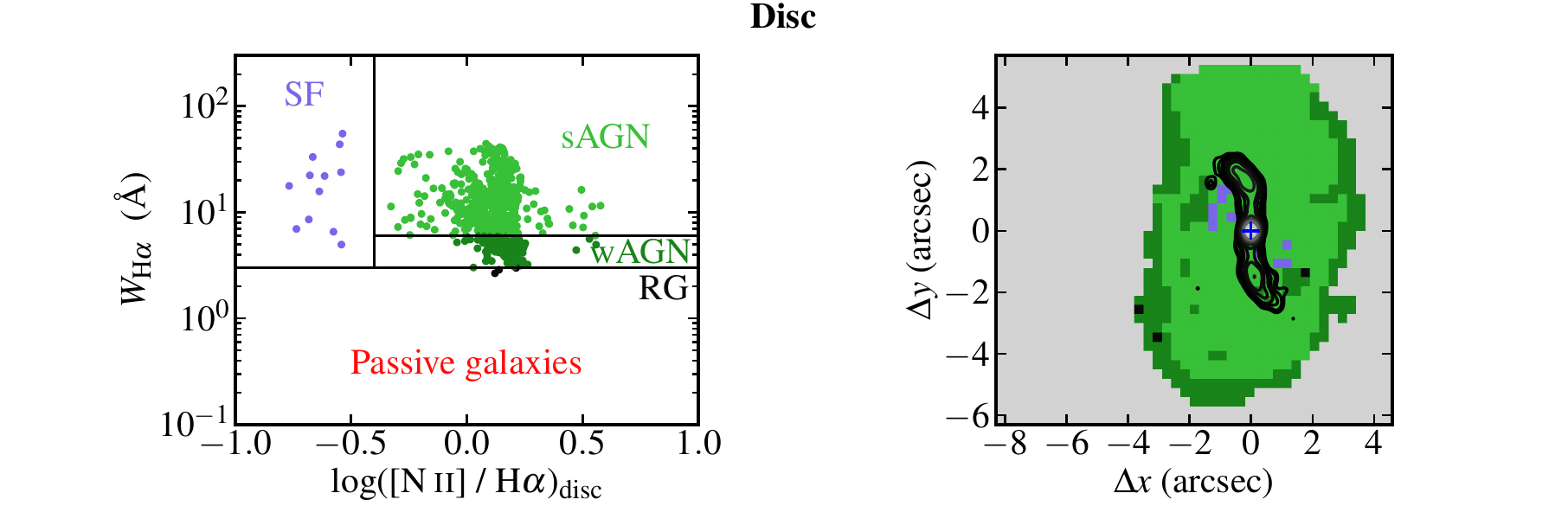}
  \includegraphics[width=18cm]{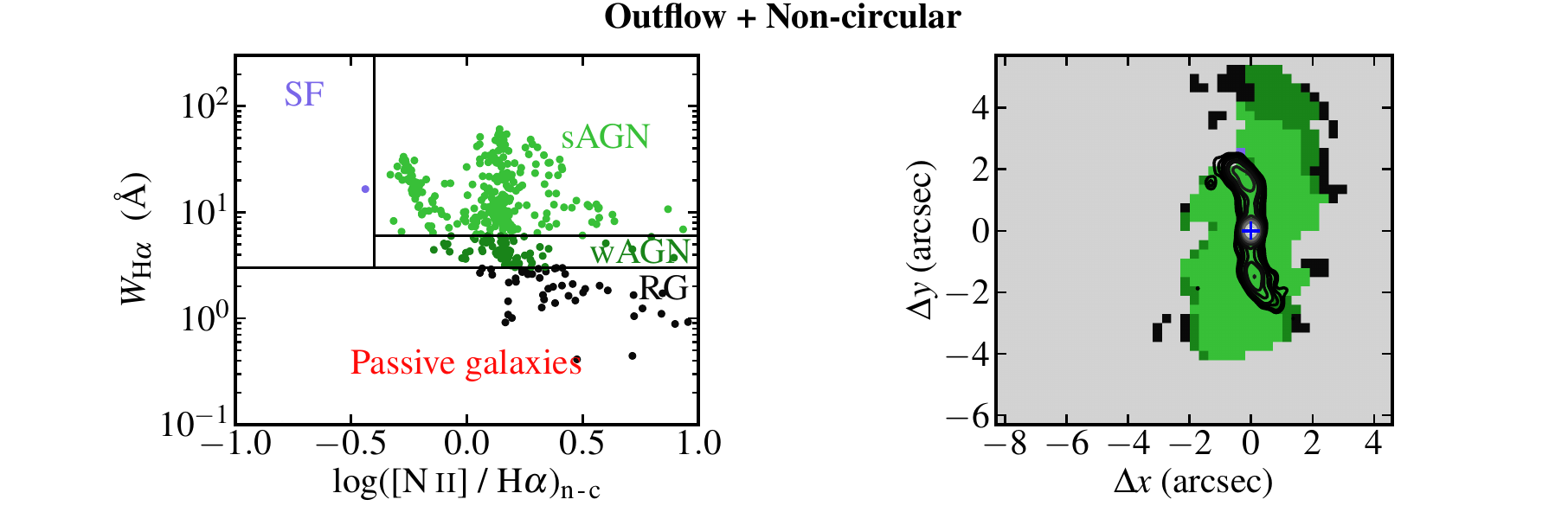}
  \caption{WHAN diagrams (left panels) for the disc
    component (top) and for the outflow and non-circular
    motion components (bottom) and the corresponding spatial
    distribution of the WHAN regions (right panels). See
    \cite{CidFernandes2011} for details for the different region
    definitions. SF: star forming galaxies, sAGN: strong AGN, wAGN:
    weak AGN, and RG: retired galaxies. }
  \label{fig:whan}
\end{figure*}

Along the galaxy minor axis, the MEGARA p-v diagram shows at the nuclear
position a broad range of velocities, both redshifted and blueshifted
velocities, with the latter reaching maximum outflow velocities in excess of
$\simeq -1000\,\mathrm{km\,s}^{-1}$ at the AGN position. We note that in the
blue part of the spectrum in the p-v diagrams the two [\ion{O}{III}] lines start
mixing as they are about $2870\,\mathrm{km\,s}^{-1}$ apart. Nevertheless, the
observed high velocities are beyond the range of velocities attributable to any
possible effect of smearing of rotation curve at the AGN position $(\simeq
400\,\mathrm{km\,s}^{-1}$). The large blueshifts observed along both axes can be
associated with the broadest of the three components fitted for the nuclear
spectrum (see Table~\ref{tab:nucleus} and Fig.~\ref{fig:spec-center-oiii}), with
$\sigma = 662\,\mathrm{km\,s}^{-1}$. Finally, there is [\ion{O}{III}] emission
in the two quadrants \emph{forbidden} by disc rotation along the major axis p-v
diagram. They show velocities of a few hundred km\,s$^{-1}$, seen as redshifted
motions to the northwest and blueshifted motions to the southeast. If we assumed
these motions are in the plane of the galaxy, they would imply
counter-rotation. However, given the velocities and velocity dispersions fitted
in the outflow region, it is more reasonable to assume that they trace
outflowing gas outside the plane of the galaxy in NGC~2110.


\subsection{Electron densities}

Figure~\ref{fig:density} shows the electron density maps for each of the two
kinematic components using the [\ion{S}{II}] method. The outflow region presents
a median value of $n_\mathrm{e} \sim$600 cm$^{-3}$, which is considerably higher
than the values of the non-circular motions and the disc. The latter has a
median value of $\sim$175 cm$^{-3}$. Analogous to our work, \citet{Kakkad2018}
studied the spatial distribution of electron density for the outflow component
of NGC~2110 (their Fig.~9). Although the comparison is not straightforward,
mainly due to their poorer seeing conditions, they found the same tendency of
higher values within the outflow region and smaller values for higher distances
of this component.

\citet{Davies2020} estimated the electron density for an aperture of $1.8\arcsec
\times 1.8\arcsec$ around the nucleus with three different methods without
splitting the flux in different kinematic components. In particular, they found
a value of 350 cm$^{-3}$ from the [\ion{S}{II}] doublet. This is in the middle
of our estimations for the two components, indicating a good agreement with
those authors. We note, however, that with the ionisation parameter method they
derived values of the electron density of up to 43000\,cm$^{-3}$, in agreement
with our estimates using the same method (see the electron density maps in
Fig.~\ref{fig:density-log_u}).

\begin{figure*} 
  \includegraphics[width=18cm]{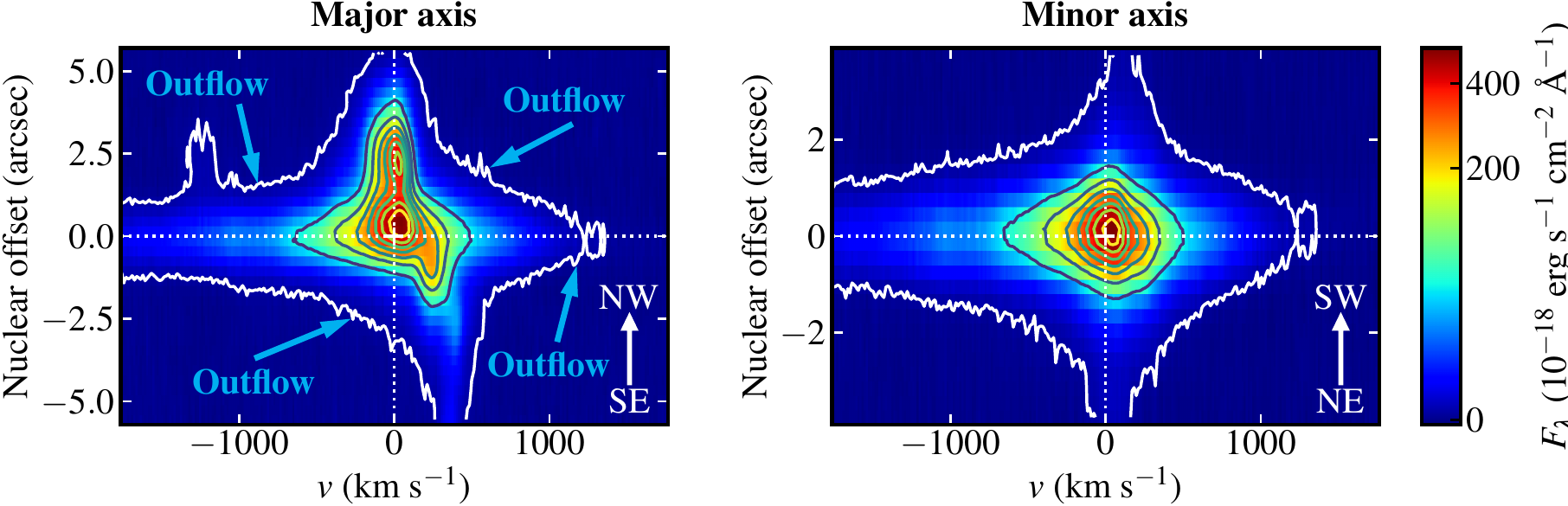}
  \caption{GTC/MEGARA p-v diagrams extracted for the [\ion{O}{III}]$\lambda$5007
    line after continuum subtraction, along the major (PA=$163^\circ$, left
    panel) and minor (PA=$73^\circ$, right panel) axes of NGC~2110. The first
    contour is plotted at three times the noise at approximately 1350\,km
    s$^{-1}$. The second contour is at five times the noise, reaching velocities
    of $-650/500\,\mathrm{km \,s}^{-1}$, and the rest of the contours are on a
    linear scale.}
  \label{fig:pv}
\end{figure*}


\subsection{Outflow properties} \label{sec:outflow}

In this section, we compute the outflow properties using the
  [\ion{O}{III}] line,  as in \cite{Davies2020},  to allow for a comparison with their work.
In the next section, we also estimate some
  outflow properties using H$\alpha$.
Before we estimated the outflowing gas mass from the [\ion{O}{III}] luminosity
(see below), we computed the extinction in the outflow region, using the
integrated H$\alpha$ and H$\beta$ fluxes. We made use of the extinction law from
\citet{Cardelli1989} with a parameter $R_V = 3.1$, assuming an intrinsic ratio
between the Balmer lines of H$\alpha$ and H$\beta$ of 3.1. We then used the
derived extinction to correct the observed [\ion{O}{III}] flux. Our median
extinction value ($A_V=0.82\,$mag) over the outflow region is lower than the
values reported by \citet{Davies2020} and \citet{Thomas2017}, which are 1.7 and
2.4\,mag, respectively. Apart from the different physical sizes used for the
spectra extraction, we also checked whether the differences are mainly due to
the fact that these works did not split the flux in kinematic components for
their estimations. Interestingly, we derived an extinction of $A_V = 2.0\,$mag
for the disc component for an aperture as that of
\citet{Davies2020}. Thus, this may
explain thus the differences. Nevertheless, we corrected the [\ion{O}{III}]
fluxes within the outflow region with the corresponding median value of the
extinction.

Considering equation B3 of \citet{Fiore2017}, the mass of ionised gas based on
the [\ion{O}{III}] luminosity can be expressed as \citep[see
  also][]{Carniani2015}:
\begin{equation}
  M^{[\ion{O}{III}]} = 4.0 \times 10^7 \mathrm{M_\odot}
  \frac{C}{10^\mathrm{[O/H]}}
  \frac{L^{[\ion{O}{III}]}}{10^{44} \ \mathrm{erg \ s^{-1}}}
  \frac{1000 \ \mathrm{cm^{-3}}}{n_\mathrm{e}}, \label{eq:mass-oiii}
\end{equation}
where $C$ is the condensation factor and [O/H] the oxygen abundance ratio (i.e.,
the ratio of AGN oxygen abundance compared to that of the Sun on a logarithmic
scale), we derived an [\ion{O}{III}] mass map. For simplicity, we assumed that
the gas clouds have the same electron density, leading to a value of $C = 1$.
$A(\mathrm{X})$ is the elemental abundance expressed as the logarithm of the
number of atoms of an element X per $10^{12}$ atoms of hydrogen. The oxygen
abundance ratio is related to a AGN oxygen abundance and that of the Sun as
follows: $\mathrm{[O/H]} = A(\mathrm{O}) - A(\mathrm{O})_\odot$, where the
required oxygen abundances for NGC~2110 and the Sun are known from the
literature: $A(\mathrm{O}) = 8.87$ \citep{Dors2015} and $A(\mathrm{O})_\odot =
8.86$ \citep{Centeno2008}.

Finally, for estimating the total outflowing mass of ionised gas $M$ in each
spaxel (using the corresponding value of the electron density), we used the
approximation adopted by \citet{Fiore2017} in terms of the [\ion{O}{III}]
outflowing mass:
\begin{equation}
  M_{\rm out} = 3 \times M^{[\ion{O}{III}]}. \label{eq:mass-total}
\end{equation}

We obtained an ionised mass in the outflow of $9.8\times 10^{4}\, M_\odot$. We
note that in the following section we provide another estimate of the outflowing
mass of ionised gas using H$\alpha$ measurements.

For the NGC~2110 outflow it seems reasonable to adopt a constant outflow history
through a cone/sphere to the outflow properties \citep[see][]{Lutz2020}. This
assumption leads to the following equations for the mass outflow rate
$\dot{M}_\mathrm{out}$, the kinetic energy $E_\mathrm{kin}$, the kinetic power
$\dot{E}_\mathrm{kin}$ and the momentum rate $\dot{P}$:
\begin{eqnarray}
  \dot{M}_\mathrm{out} & = & M_\mathrm{out} \frac{v_\mathrm{out}}{r_\mathrm{out}}
    \label{eq:mass} \\
  E_\mathrm{kin}       & = & \frac{1}{2} M_\mathrm{out} v_\mathrm{out}^2
    \label{eq:kin-energy} \\
  \dot{E}_\mathrm{kin} & = & \frac{1}{2} \dot{M}_\mathrm{out} v_\mathrm{out}^2
    \label{eq:kin-power} \\
  \dot{P}             & = & \dot{M}_\mathrm{out} v_\mathrm{out}
    \label{eq:momentum-rate}
\end{eqnarray}
where $M_\mathrm{out}$ is the ionised gas mass integrated within the outflow
region, $v_\mathrm{out}$ the outflow velocity, and $r_\mathrm{out}$ the outflow
radius.

In order to estimate the outflow velocity $v_\mathrm{out}$, we used a similar
approach to that of \citet{Davies2020}. We computed the percentiles 2 and 98 of
the velocity distribution inferred from the outflow profile of the
[\ion{O}{III}] doublet, measured their differences with respect to the systemic
velocity, and chose whichever is larger. We obtained a maximum outflowing
velocity of $v_{98} = 1167\,\mathrm{km \,s}^{-1}$, whereas they reported a value
of $1665\,\mathrm{km \,s}^{-1}$. We note that apart from the different
extraction sizes, they assumed that the whole profile is dominated by the
outflow, while we built the outflow profile stacking the line profiles of the
non-circular component in each spaxel within the outflow region. As a note, our
kinematic decomposition allowed us to estimate that approximately half of the
flux of the whole profile in the $\sim 4\arcsec$ region comes from the disc
component.

For the outflow radius we took $r_\mathrm{out}  =
400 \ \mathrm{pc}$ (projected value, see Fig.~\ref{fig:flux-blue}), which is 2.7 times 
larger than the value adopted by \citet{Davies2020}. The outflow region size is
approximately consistent with two times the measured FWHM at the wavelength of
[\ion{O}{III}] (see Sect.~\ref{subsec:outflowmorphology}). We also checked that
our estimation is equivalent to that employed by \citet{Kang2018}, who defined
the outflow radius as the distance in which the velocity dispersion from the
[\ion{O}{III}] doublet changes with respect to the stellar values. Indeed, the
observed velocity dispersions in the outflow region of NGC~2110 (see e.g.,
Fig.~\ref{fig:sigma-blue}) are higher than the stellar velocity dispersion
\citep[$220-260\,$km s$^{-1}$,
  see][]{Nelson1995,GonzalezDelgado2002,Burtscher2021}. However, the measured
radius for NGC~2110 falls approximately 4$\sigma$ below the expected value from
the correlation between the outflow size and [\ion{O}{III}] luminosity reported
by \citet{Kang2018} for more luminous AGN. Table~\ref{tab:outflow} summarises the properties of the outflow.

\begin{table}
  \caption{Outflow properties from the [\ion{O}{III}] line.}
  \label{tab:outflow}
  \centering
  \begin{tabular}{l c}
    \hline\hline
    Property & Value\\
    \hline
    $M_\mathrm{out} \ (\mathrm{M_\odot})$ &
    $9.8 \times 10^{4}$ \\
    $\dot{M}_\mathrm{out} \ (\mathrm{M_\odot \ yr^{-1}})$ &
    $0.29$\\
    $E_\mathrm{kin} \ (\mathrm{erg})$ &
    $1.3 \times 10^{54}$ \\
    $\dot{E}_\mathrm{kin} \ (\mathrm{erg \ s^{-1}})$ &
    $1.3 \times 10^{41}$ \\
    $\dot{P} \ (\mathrm{dyn})$ &
    $2.2 \times 10^{33}$ \\
    \hline
  \end{tabular}
\end{table}

\citet{Davies2020} also computed the outflow properties of NGC~2110 and derived
a mass outflow rate that is approximately an order magnitude lower. Their
$L^{[\ion{O}{III}]}$ is 0.12 dex lower than ours. Their lower flux calibration
(by 0.40 dex, see Sect.~\ref{subsec:optical_data}) is compensated with
their extinction correction, which is 0.39 dex higher. Therefore, the difference
in luminosity is due to their light loss caused by their smaller aperture (0.20
dex), even though they assumed that all the light can be attributed to the
outflow. Regarding $M_\mathrm{out}$, they obtained a value 1.59 dex lower, which
is mainly due to their much higher value of the electron density. For the
ionised gas mass rate $\dot{M}_\mathrm{out}$, the difference decreases to 0.99
dex because we considered an outflowing velocity 0.15 dex lower and a radius
0.43 higher (which in both cases help to alleviate the difference). The same
happens for the kinetic power $\dot{E}_\mathrm{kin}$. All this highlights the
importance of understanding the assumptions for estimating the outflow
properties while comparing with results from other works.


\subsection{Are the outflow properties linked with the radio jet?}
\label{subsec:outflow-jet}

NGC~2110 follows well the scaling relation of the ionised
mass outflow rate and the 
AGN bolometric luminosity derived by \cite{Fiore2017} for luminous AGN
using  $\log L_\mathrm{AGN} = 44.5\,\mathrm{erg\,s}^{-1}$
from \cite{Davies2020}. However, it falls below the
scaling relation with the kinetic power. In 
this respect, NGC~2110 behaves like other AGN of similar luminosities
\citep[see e.g., Fig.~10 of][]{GarciaBernete2021}, which show a large
range of kinetic powers. Moreover, NGC~2110
lies in the upper envelop of this observed large scatter at a given AGN luminosity. It is possible that, at least in
cases such as NGC~2110,
some of the kinetic power, in addition to that provided by the AGN
radiation pressure, might be due to the effect of a moderate power
radio jet. In this section we explore this scenario in NGC~2110.

\citet{Venturi2021} observed in a small sample of Seyfert galaxies that regions
with high velocity dispersion were generally perpendicular to the orientation of
the radio jet. In NGC~2110, the increased velocity dispersions,
compared to those of the disc, are observed both perpendicular and parallel to
the radio jet. Equally conspicuous is the fact that the bending of the radio
jet, to the northeast and southwest within the outflow region of NGC~2110,
coincides with the sharp transition in velocity dispersions seen in the region
with the non-circular motions (right panels of Figs.~\ref{fig:sigma-blue} and
\ref{fig:sigma-red}). The length of the linear part of the radio jet nearly
matches that of the outflow region. In this this region, there is also weak
CO(2-1) emission, termed the "lacuna", along the northwest-southeast direction
(PA$\simeq -25^{\circ}$) passing through the AGN location
\citep{Rosario2019}. In this section, we investigate whether the outflow
properties of NGC~2110 follow the relations between the jet radio power and the
outflowing ionised mass and kinetic energy found by \cite{Venturi2021}.

\begin{figure*}
  \centering
  \includegraphics{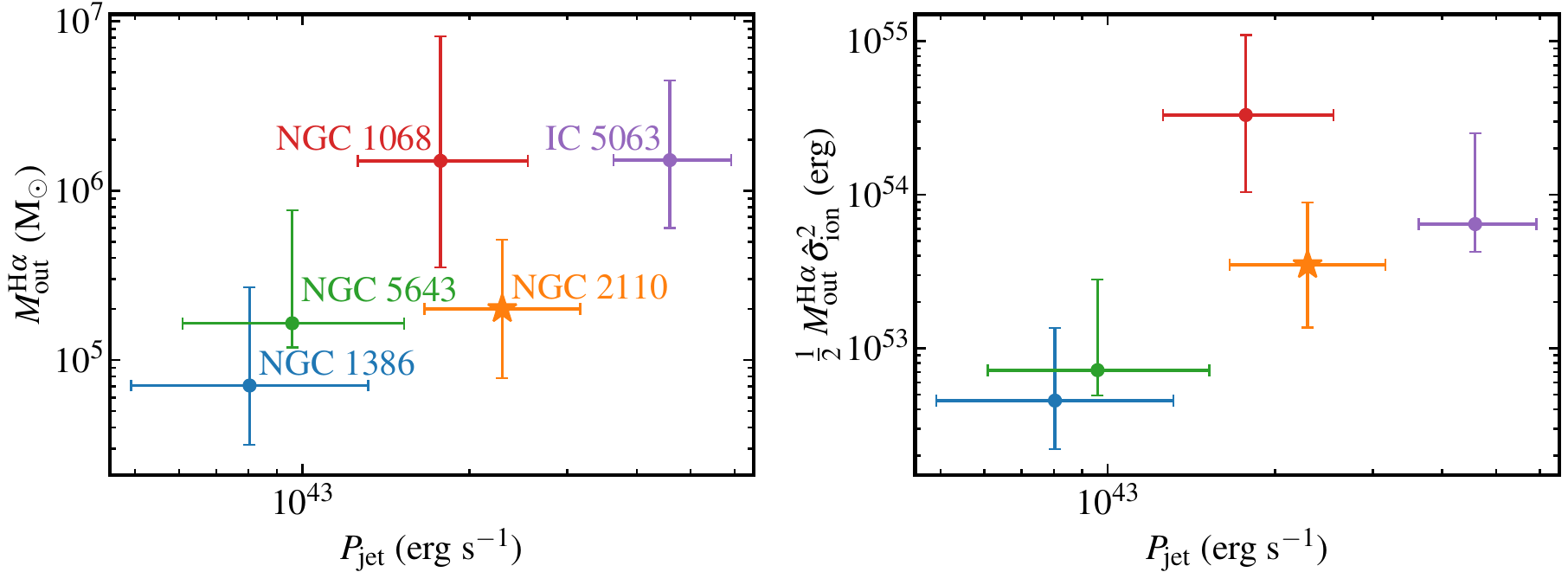}
  \caption{Outflow mass (left panel) and a proxy for the kinetic energy (right
    panel) versus the jet radio power. Circles are galaxies from
    \citet{Venturi2021}, while the star corresponds to NGC~2110 (see
    Sect.~\ref{subsec:outflow-jet}). For this galaxy, the outflowing ionised gas
    mass corresponds only to that observed in the outflow region. The horizontal
    error bars for NGC~2110 take into account the errors in equation 16 of
    \citet{Birzan2008}, while vertical error bars include the uncertainties
    related to the extinction correction and the relative flux calibration
    between the two MEGARA gratings, as well as the uncertainty from the
    absolute flux calibration.}
  \label{fig:jet-power}
\end{figure*}

We estimated the kinetic power of the radio jet for NGC~2110 from the 20\,cm
radio flux reported by \cite{Nagar1999}. We converted the total radio flux into
kinetic power using equation 16 of \citet{Birzan2008}, as done by
  \cite{Venturi2021}. We note, however, that other calibrations might
  provide kinetic powers differing by up to an order of magnitude. We obtained
$P_\mathrm{jet} = 2.3 \times 10^{43} \ \mathrm{erg\,s^{-1}}$ for NGC~2110, which is slightly
higher than that of NGC~1068 and approximately a factor of two less powerful
than the jet in IC~5063 \citep{Venturi2021}.

To make a proper comparison with \citet{Venturi2021}, we estimated the
outflowing mass with our extinction corrected H$\alpha$ luminosity and their
equation 1, instead of our equations \ref{eq:mass-oiii} and
\ref{eq:mass-total}. Unlike their work, for NGC~2110 we only included the mass
from the outflow component. With this approach, the outflowing ionised mass
within the outflow region is $M_\mathrm{out}^\mathrm{H\alpha} = 2.0 \times
10^{5} \mathrm{M_\odot}$, which is a factor of two higher than our value
estimated from the [\ion{O}{III}] line (Table~\ref{tab:outflow}). Finally, we
would like to emphasise that our mass estimates and theirs are based on electron
densities from the [\ion{S}{II}] method. Following \cite{Venturi2021}, the
kinetic energy of the jet is $M_\mathrm{out}\hat{\sigma} _\mathrm{ion}^2/2$. We
measured the velocity dispersion of the [\ion{O}{III}] profile including both
components, as in their work, that is from its second-order moment. We obtained
$\hat{\sigma}_\mathrm{ion}$=419 km s$^{-1}$.

In Fig.~\ref{fig:jet-power} we compare NGC~2110 with the trends observed by
\citet{Venturi2021} between the outflow properties and the jet radio power.
NGC~2110 appears to follow well the trends suggested by their galaxies, except
for the more extreme properties of NGC~1068. The derived kinetic energy in
NGC~2110 is between NGC~5643 and IC~5063, which are well-known for hosting a
radio jet that is close to the disc of the galaxy and driving a multi-phase
energetic outflow \cite[see][and references
  therein]{Morganti2015,Venturi2021}. On the other hand, in
these diagrams NGC~1068 is well above the trends. As in NGC~2110, the radio jet
of NGC~1068 suffers a reorientation due probably to the impact with a gas cloud
in the disc of the galaxy, and further away (a few hundred pc from the AGN) the
radio morphology resembles a bow-shock
\citep{GarciaBurillo2014,GarciaBurillo2017}. However unlike NGC~2110, in
NGC~1068 shocks only dominate the gas excitation at the radio jet location,
while AGN and star formation excitation dominate in other regions
\citep{DAgostino2019}.

Considering the above observational evidence, we speculate that the radio jet
propagated through the nuclear region of NGC~2110 and shocked the gas producing
highly turbulent motions, both parallel and perpendicular to the radio
jet. Indeed, \cite{Ulvestad1983} proposed that that the radio jet morphology of
NGC~2110 is due to ram pressure bending by a dense rotating disc. Simulations
predict that the gas would expand like a bubble
\citep[e.g.,][]{Wagner2011,Mukherjee2018}. Since in NGC~2110 the jet is not
perpendicular to the galaxy disc \citep{Pringle1999}, the jet-inflated bubble
probably ended up impacting the ISM of the galaxy disc. As a result the jet was
deflected and probably decelerated. This may have caused a decrease in the
turbulence of the gas, and thus lower observed velocity dispersions in the
non-circular component outside the outflow region. In the disc of the galaxy we
observe shock-like emission in the central region where the jet is still
operating but at distances beyond the radio bend the dominance of shocks
diminishes and it is then possible to identify gas excited by the AGN (see
Fig~\ref{fig:bpt-disc}).

 In summary, the discussion above provides additional support for the
scenario  where part of the observed ionised gas outflow in NGC~2110 could be
  driven by the radio jet when it expands laterally and radially in
  the disc of the galaxy.


\section{Summary and conclusions} \label{sec:conclusions}

In this paper, we presented new GTC/MEGARA IFU observations of the central $\sim
13\arcsec \times 11\arcsec$ ($\sim 2\,\mathrm{kpc} \times 1.8\,\mathrm{kpc}$ in
projection) region of NGC~2110 with $R\simeq 5000-5900$. We used two gratings to
observe the most prominent nebular emission lines in the $4300- 5200\,\AA$ and
$6100-7300\,\AA$ spectral ranges. The goal was to study the spatially-resolved
properties of the ionised gas in the central 2\,kpc of NGC~2110, including its
outflow, and evaluate the role of its radio jet. We fitted the emission lines
with a maximum of two Gaussian components, which was sufficient in the majority
of spaxels, except at the AGN position. We performed a kinematic separation
using the observed velocities and velocity dispersions and a comparison with the
stellar kinematics from \cite{GonzalezDelgado2002} and \cite{Ferruit2004}.

We identified the ionised gas rotation in the disc of the galaxy with an
intermediate velocity dispersion component ($\sigma \simeq 60 -200\,\mathrm{km\,
  s}^{-1}$). It is detected in all the bright emission lines ([\ion{O}{III}],
H$\alpha$, [\ion{N}{II}] and [\ion{S}{II}]) over a $\sim 8 \arcsec \times 12
\arcsec$ ($1.3 \times 1.9$ kpc$^2$ in projection), which corresponds to most of
the MEGARA FoV along the major axis of the galaxy.

The non-circular component covers a $\sim4\arcsec \times 8\arcsec$ area and
presents different properties in two distinct regions. The first is a
spatially-resolved region, mostly in the north-south direction, with a total
size of $\sim 2.5\arcsec \sim 400\,$pc around the nucleus. It presents the
highest velocity dispersions, typically 700\,km s$^{-1}$, and is identified with
the ionised outflow in this galaxy. The extent of the outflow region, as traced
by the [\ion{O}{III}] line, is approximately equal to the linear part of the
radio jet. Outside the outflow region, beyond the edges of the radio source and
where it is deflected, the kinematics of the gas with non-circular motions
changes drastically. The velocity dispersion drops to $<60\,$km s$^{-1}$. These
are likely non- circular motions taking place in the disc of the galaxy,
possibly unrelated to the inner outflow and the radio jet.

We fitted the AGN nuclear spectrum of NGC~2110 with three Gaussian components,
with the broadest [\ion{O}{III}] component reaching a velocity dispersion of
662\,km\,s$^{-1}$. The [\ion{O}{III}] p-v diagrams along the major and minor
axes of the galaxy reveal at the AGN position observed velocities in excess of
$1000\,\mathrm{km\,s}^{-1}$, both blueshifted and redshifted. There are other
redshifted and blueshifted components of several hundreds of km\,s$^{-1}$ in the
p-v diagram along the major axis of the galaxy. Their location in the two
quadrants \emph{forbidden} by rotation indicates that part of the outflowing gas
is outside the disc of the galaxy.

The spatially-resolved BPT diagnostic diagrams of the outflow region reveal
mostly LI(N)ER-like excitation. The observed line ratios are consistent with
predictions from shock models. In the disc of the galaxy, there is a large
number of spaxels with LI(N)ER-like excitation, but also `Seyfert-like'
excitation due to AGN photoionisation. The latter excitation appears in regions
beyond, although close to, the region where the radio jet is bent, at projected
distances greater than approximately $1-2\arcsec$, reaching at least $5\arcsec
\simeq 800\,$pc. This is likely gas in the host galaxy disc being illuminated by
AGN photons at the intersection of a ionisation cone with the galaxy disc
\citep[see also][]{Rosario2010}.

Using the outflow component of the [\ion{O}{III}]$\lambda$5007 line, we derived
an outflowing ionised gas mass, ionised mass outflow rate, and outflow kinetic
power of $M_\mathrm{out} = 9.8 \times 10^{4} \ \mathrm{M_\odot}$,
$\dot{M}_\mathrm{out} = 0.29 \ \mathrm{M_\odot \ yr^{-1}}$ and
$\dot{E}_\mathrm{kin} = 1.3 \times 10^{41} \ \mathrm{erg \ s^{-1}}$,
respectively. As found in other Seyfert galaxies, NGC~2110 lies
  below the correlation between the AGN bolometric luminosity and kinetic power
  of the ionised outflow derived for luminous AGN. However, NGC~2110 shows an
  increased kinetic power when compared to other Seyferts of similar
  AGN luminosity. We explored the
  possibility that part of the 
 kinetic power might be due to the effect of the moderate power radio jet
 detected in this galaxy. To do so, we  compared NGC~2110 
  outflowing gas mass and kinetic power properties  with
its jet radio power.   NGC~2110 follows well the 
observational trends between these quantities  found by \citet{Venturi2021}  for a few
  Seyfert galaxies.  We proposed  that
the ionised gas outflow in the inner $\simeq 400\,$pc region of NGC~2110 is driven
in part by the radio jet. This is supported by the presence of
shock-like emission, presumably induced by the passage of the moderate
luminosity radio jet through the disc. However, beyond the outflow region there
is also gas photoionised by the AGN in the central 2\,kpc of NGC~2110. This gas
is rotating and located in regions bordering and past the radio jet beding,
where it is being illuminated by AGN photons.


\begin{acknowledgements}
  We would like to thank the referee for insightful comments and
   O. Gonz\'alez-Mart\'{\i}n for providing constructive
  feedback. LPdA, AAH, SGB and MVM acknowledge financial support from grant
  PGC2018-094671-B-I00 funded by MCIN/AEI/10.13039/501100011033 and by
  ERDF A way of making Europe. AAH and MVM also acknowledge financial
  support from grant PID2021-124665NB-I00 funded by the Spanish Ministry of
  Science and Innovation and the State Agency of Research MCIN/AEI/
  10.13039/501100011033 and ERDF A way of making Europe.
  SGB acknowledges support from the research project
  PID2019-106027GA-C44 of the Spanish Ministerio de Ciencia e Innovaci\'on. IGB
  and DR acknowledge support from STFC through grant ST/S000488/1. BGL
  acknowledges support from grants PID2019-107010GB-100 and the Severo Ochoa
  CEX2019-000920-S. CRA acknowledges the project ``Feeding and feedback in
  active galaxies'', with reference PID2019-106027GB-C42, funded by
  MICINN-AEI/10.13039/501100011033, and the European Union’s Horizon 2020
  research and innovation programme under Marie Sk\l odowska-Curie grant
  agreement No 860744 (BID4BEST). CRA and AA acknowledge the project
  ``Quantifying the impact of quasar feedback on galaxy evolution'', with
  reference EUR2020-112266, funded by MICINN-AEI/10.13039/501100011033 and the
  European Union NextGenerationEU/PRTR; the Consejer\'{\i}a de Econom\'{\i}a,
  Conocimiento y Empleo del Gobierno de Canarias and the European Regional
  Development Fund (ERDF) under grant ``Quasar feedback and molecular gas
  reservoirs'', with reference ProID2020010105, ACCISI/FEDER, UE. EB
  acknowledges the Mar\'{\i}a Zambrano program of the Spanish Ministerio de
  Universidades funded by the Next Generation European Union and is also partly
  supported by grant RTI2018-096188-B-I00 funded by
  MCIN/AEI/10.13039/501100011033. CR acknowledges support from the Fondecyt
  Iniciacion grant 11190831 and ANID BASAL project FB210003. DR acknowledges
  support from University of Oxford John Fell Fund.
  
  Based on observations made with the Gran Telescopio Canarias (GTC), installed
  in the Spanish Observatorio del Roque de los Muchachos of the Instituto de
  Astrof\'{\i}sica de Canarias, in the island of La Palma. This work is based on
  data obtained with MEGARA instrument, funded by European Regional Development
  Funds (ERDF), through Programa Operativo Canarias FEDER 2014-2020.

  This research made use of NumPy \citep{Harris2020}, SciPy
  \citep{Virtanen2020}, Matplotlib \citep{Hunter2007} and Astropy
  \citep{Astropy2013,Astropy2018}.

  This research has made use of the NASA/IPAC Extragalactic Database (NED),
  which is funded by the National Aeronautics and Space Administration and
  operated by the California Institute of Technology.
  
  This research has made use of ESASky, developed by the ESAC Science Data
  Centre (ESDC) team and maintained alongside other ESA science mission's
  archives at ESA's European Space Astronomy Centre (ESAC, Madrid, Spain).
\end{acknowledgements}


\bibliographystyle{aa}
\bibliography{bibliography}

\begin{thebibliography}{83}
\expandafter\ifx\csname natexlab\endcsname\relax\def\natexlab#1{#1}\fi

\bibitem[{{Astropy Collaboration} {et~al.}(2018){Astropy Collaboration},
  Price-Whelan, Sip{\H{o}}cz, G{\"u}nther, Lim, Crawford, Conseil, Shupe,
  Craig, Dencheva, Ginsburg, VanderPlas, Bradley, P{\'e}rez-Su{\'a}rez,
  de~Val-Borro, Aldcroft, Cruz, Robitaille, Tollerud, Ardelean, Babej, Bach,
  Bachetti, Bakanov, Bamford, Barentsen, Barmby, Baumbach, Berry, Biscani,
  Boquien, Bostroem, Bouma, Brammer, Bray, Breytenbach, Buddelmeijer, Burke,
  Calderone, Cano~Rodr{\'\i}guez, Cara, Cardoso, Cheedella, Copin, Corrales,
  Crichton, D'Avella, Deil, Depagne, Dietrich, Donath, Droettboom, Earl, Erben,
  Fabbro, Ferreira, Finethy, Fox, Garrison, Gibbons, Goldstein, Gommers, Greco,
  Greenfield, Groener, Grollier, Hagen, Hirst, Homeier, Horton, Hosseinzadeh,
  Hu, Hunkeler, Ivezi{\'c}, Jain, Jenness, Kanarek, Kendrew, Kern, Kerzendorf,
  Khvalko, King, Kirkby, Kulkarni, Kumar, Lee, Lenz, Littlefair, Ma, Macleod,
  Mastropietro, McCully, Montagnac, Morris, Mueller, Mumford, Muna, Murphy,
  Nelson, Nguyen, Ninan, N{\"o}the, Ogaz, Oh, Parejko, Parley, Pascual, Patil,
  Patil, Plunkett, Prochaska, Rastogi, Reddy~Janga, Sabater, Sakurikar,
  Seifert, Sherbert, Sherwood-Taylor, Shih, Sick, Silbiger, Singanamalla,
  Singer, Sladen, Sooley, Sornarajah, Streicher, Teuben, Thomas, Tremblay,
  Turner, Terr{\'o}n, van Kerkwijk, de~la Vega, Watkins, Weaver, Whitmore,
  Woillez, Zabalza, \& Contributors}]{Astropy2018}
{Astropy Collaboration}, Price-Whelan, A.~M., Sip{\H{o}}cz, B.~M., {et~al.}
  2018, \aj, 156, 123

\bibitem[{{Astropy Collaboration} {et~al.}(2013){Astropy Collaboration},
  Robitaille, Tollerud, Greenfield, Droettboom, Bray, Aldcroft, Davis,
  Ginsburg, Price-Whelan, Kerzendorf, Conley, Crighton, Barbary, Muna,
  Ferguson, Grollier, Parikh, Nair, Unther, Deil, Woillez, Conseil, Kramer,
  Turner, Singer, Fox, Weaver, Zabalza, Edwards, Azalee~Bostroem, Burke, Casey,
  Crawford, Dencheva, Ely, Jenness, Labrie, Lim, Pierfederici, Pontzen, Ptak,
  Refsdal, Servillat, \& Streicher}]{Astropy2013}
{Astropy Collaboration}, Robitaille, T.~P., Tollerud, E.~J., {et~al.} 2013,
  \aap, 558, A33

\bibitem[{{Audibert} {et~al.}(2023){Audibert}, {Ramos Almeida},
  {Garc{\'\i}a-Burillo}, {Combes}, {Bischetti}, {Meenakshi}, {Mukherjee},
  {Bicknell}, \& {Wagner}}]{Audibert2023}
{Audibert}, A., {Ramos Almeida}, C., {Garc{\'\i}a-Burillo}, S., {et~al.} 2023,
  \aap, 671, L12

\bibitem[{{Baines} {et~al.}(2017){Baines}, {Giordano}, {Racero}, {Salgado},
  {L{\'o}pez Mart{\'\i}}, {Mer{\'\i}n}, {Sarmiento}, {Guti{\'e}rrez}, {Ortiz de
  Landaluce}, {Le{\'o}n}, {de Teodoro}, {Gonz{\'a}lez}, {Nieto}, {Segovia},
  {Pollock}, {Rosa}, {Arviset}, {Lennon}, {O'Mullane}, \& {de
  Marchi}}]{Baines2017}
{Baines}, D., {Giordano}, F., {Racero}, E., {et~al.} 2017, \pasp, 129, 028001

\bibitem[{Baldwin {et~al.}(1981)Baldwin, Phillips, \& Terlevich}]{Baldwin1981}
Baldwin, J.~A., Phillips, M.~M., \& Terlevich, R. 1981, \pasp, 93, 5

\bibitem[{Baron \& Netzer(2019)}]{Baron2019}
Baron, D. \& Netzer, H. 2019, \mnras, 486, 4290

\bibitem[{{Bellocchi} {et~al.}(2019){Bellocchi}, {Ascasibar}, {Galbany},
  {S{\'a}nchez}, {Ibarra-Medel}, {Gavil{\'a}n}, \& {D{\'\i}az}}]{Bellocchi2019}
{Bellocchi}, E., {Ascasibar}, Y., {Galbany}, L., {et~al.} 2019, \aap, 625, A83

\bibitem[{B{\^\i}rzan {et~al.}(2008)B{\^\i}rzan, McNamara, Nulsen, Carilli, \&
  Wise}]{Birzan2008}
B{\^\i}rzan, L., McNamara, B.~R., Nulsen, P. E.~J., Carilli, C.~L., \& Wise,
  M.~W. 2008, \apj, 686, 859

\bibitem[{{Bremer} {et~al.}(2013){Bremer}, {Scharw{\"a}chter}, {Eckart},
  {Valencia-S.}, {Zuther}, {Combes}, {Garcia-Burillo}, \&
  {Fischer}}]{Bremer2013}
{Bremer}, M., {Scharw{\"a}chter}, J., {Eckart}, A., {et~al.} 2013, \aap, 558,
  A34

\bibitem[{Burtscher {et~al.}(2021)Burtscher, Davies, Shimizu, Riffel, Rosario,
  Hicks, Lin, Riffel, Schartmann, Schnorr-M{\"u}ller, Storchi-Bergmann,
  Orban~de Xivry, \& Veilleux}]{Burtscher2021}
Burtscher, L., Davies, R.~I., Shimizu, T.~T., {et~al.} 2021, \aap, 654, A132

\bibitem[{Cardelli {et~al.}(1989)Cardelli, Clayton, \& Mathis}]{Cardelli1989}
Cardelli, J.~A., Clayton, G.~C., \& Mathis, J.~S. 1989, \apj, 345, 245

\bibitem[{{Carniani} {et~al.}(2015){Carniani}, {Marconi}, {Maiolino},
  {Balmaverde}, {Brusa}, {Cano-D{\'\i}az}, {Cicone}, {Comastri}, {Cresci},
  {Fiore}, {Feruglio}, {La Franca}, {Mainieri}, {Mannucci}, {Nagao}, {Netzer},
  {Piconcelli}, {Risaliti}, {Schneider}, \& {Shemmer}}]{Carniani2015}
{Carniani}, S., {Marconi}, A., {Maiolino}, R., {et~al.} 2015, \aap, 580, A102

\bibitem[{Carrasco {et~al.}(2018)Carrasco, Gil~de Paz, Gallego,
  Iglesias-P{\'a}ramo, Cedazo, Garc{\'\i}a~Vargas, Arrillaga, Avil{\'e}s,
  Bouquin, Carbajo, Cardiel, Carrera, Castillo~Morales,
  Castillo-Dom{\'\i}nguez, Esteban San~Rom{\'a}n, Ferrusca,
  G{\'o}mez-{\'A}lvarez, Izazaga-P{\'e}rez, Lefort, L{\'o}pez~Orozco,
  Maldonado, Mart{\'\i}nez~Delgado, Morales~Dur{\'a}n, M{\'u}jica, Ortiz,
  P{\'a}ez, Pascual, P{\'e}rez-Calpena, Picazo, S{\'a}nchez-Penim,
  S{\'a}nchez-Blanco, Tulloch, Vel{\'a}zquez, V{\'\i}lchez, Zamorano, Aguerri,
  Barrado, Bertone, Cava, Catal{\'a}n-Torrecilla, Cenarro, Ch{\'a}vez, Dullo,
  Eliche, Garc{\'\i}a, Garc{\'\i}a-Rojas, Guichard, Gonz{\'a}lez-Delgado,
  Guzm{\'a}n, Herrero, Hu{\'e}lamo, Hughes, Jim{\'e}nez-Vicente, Kehrig,
  Marino, M{\'a}rquez, Masegosa, Mayya, M{\'e}ndez-Abreu, Moll{\'a},
  Mu{\~n}oz-Tu{\~n}{\'o}n, Peimbert, P{\'e}rez-Gonz{\'a}lez, P{\'e}rez-Montero,
  Roca-F{\`a}brega, Rodr{\'\i}guez, Rodr{\'\i}guez-Espinosa,
  Rodr{\'\i}guez-Merino, Rodr{\'\i}guez-Mu{\~n}oz, Rosa-Gonz{\'a}lez,
  S{\'a}nchez-Almeida, S{\'a}nchez~Contreras, S{\'a}nchez-Bl{\'a}zquez,
  S{\'a}nchez, Sarajedini, Silich, Sim{\'o}n-D{\'\i}az, Tenorio-Tagle,
  Terlevich, Terlevich, Torres-Peimbert, Trujillo, Tsamis, \&
  Vega}]{Carrasco2018}
Carrasco, E., Gil~de Paz, A., Gallego, J., {et~al.} 2018, in Society of
  Photo-Optical Instrumentation Engineers (SPIE) Conference Series, Vol. 10702,
  Ground-based and Airborne Instrumentation for Astronomy VII, ed. C.~J.
  {Evans}, L.~{Simard}, \& H.~{Takami}, 1070216

\bibitem[{{Cazzoli} {et~al.}(2022){Cazzoli}, {Hermosa Mu{\~n}oz},
  {M{\'a}rquez}, {Masegosa}, {Castillo-Morales}, {Gil de Paz},
  {Hern{\'a}ndez-Garc{\'\i}a}, {La Franca}, \& {Ramos Almeida}}]{Cazzoli2022}
{Cazzoli}, S., {Hermosa Mu{\~n}oz}, L., {M{\'a}rquez}, I., {et~al.} 2022, \aap,
  664, A135

\bibitem[{Centeno \& Socas-Navarro(2008)}]{Centeno2008}
Centeno, R. \& Socas-Navarro, H. 2008, \apjl, 682, L61

\bibitem[{{Cid Fernandes} {et~al.}(2011){Cid Fernandes}, {Stasi{\'n}ska},
  {Mateus}, \& {Vale Asari}}]{CidFernandes2011}
{Cid Fernandes}, R., {Stasi{\'n}ska}, G., {Mateus}, A., \& {Vale Asari}, N.
  2011, \mnras, 413, 1687

\bibitem[{{Cid Fernandes} {et~al.}(2010){Cid Fernandes}, {Stasi{\'n}ska},
  {Schlickmann}, {Mateus}, {Vale Asari}, {Schoenell}, \&
  {Sodr{\'e}}}]{CidFernandes2010}
{Cid Fernandes}, R., {Stasi{\'n}ska}, G., {Schlickmann}, M.~S., {et~al.} 2010,
  \mnras, 403, 1036

\bibitem[{Clements(1983)}]{Clements1983}
Clements, E.~D. 1983, \mnras, 204, 811

\bibitem[{Combes {et~al.}(2013)Combes, Garc{\'\i}a-Burillo, Casasola, Hunt,
  Krips, Baker, Boone, Eckart, Marquez, Neri, Schinnerer, \&
  Tacconi}]{Combes2013}
Combes, F., Garc{\'\i}a-Burillo, S., Casasola, V., {et~al.} 2013, \aap, 558,
  A124

\bibitem[{Comer{\'o}n {et~al.}(2021)Comer{\'o}n, Knapen, Ramos~Almeida, \&
  Watkins}]{Comeron2021}
Comer{\'o}n, S., Knapen, J.~H., Ramos~Almeida, C., \& Watkins, A.~E. 2021,
  \aap, 645, A130

\bibitem[{{Cresci} {et~al.}(2015){Cresci}, {Marconi}, {Zibetti}, {Risaliti},
  {Carniani}, {Mannucci}, {Gallazzi}, {Maiolino}, {Balmaverde}, {Brusa},
  {Capetti}, {Cicone}, {Feruglio}, {Bland-Hawthorn}, {Nagao}, {Oliva},
  {Salvato}, {Sani}, {Tozzi}, {Urrutia}, \& {Venturi}}]{Cresci2015}
{Cresci}, G., {Marconi}, A., {Zibetti}, S., {et~al.} 2015, \aap, 582, A63

\bibitem[{{D'Agostino} {et~al.}(2019){D'Agostino}, {Kewley}, {Groves},
  {Medling}, {Di Teodoro}, {Dopita}, {Thomas}, {Sutherland}, \&
  {Garcia-Burillo}}]{DAgostino2019}
{D'Agostino}, J.~J., {Kewley}, L.~J., {Groves}, B.~A., {et~al.} 2019, \mnras,
  487, 4153

\bibitem[{Davies {et~al.}(2020)Davies, Baron, Shimizu, Netzer, Burtscher,
  de~Zeeuw, Genzel, Hicks, Koss, Lin, Lutz, Maciejewski,
  M{\"u}ller-S{\'a}nchez, Orban~de Xivry, Ricci, Riffel, Riffel, Rosario,
  Schartmann, Schnorr-M{\"u}ller, Shangguan, Sternberg, Sturm,
  Storchi-Bergmann, Tacconi, \& Veilleux}]{Davies2020}
Davies, R., Baron, D., Shimizu, T., {et~al.} 2020, \mnras, 498, 4150

\bibitem[{Davies {et~al.}(2016)Davies, Dopita, Kewley, Groves, Sutherland,
  Hampton, Shastri, Kharb, Bhatt, Scharw{\"a}chter, Jin, Banfield, Zaw, James,
  Juneau, \& Srivastava}]{Davies2016}
Davies, R.~L., Dopita, M.~A., Kewley, L., {et~al.} 2016, \apj, 824, 50

\bibitem[{Dors {et~al.}(2015)Dors, Cardaci, H{\"a}gele, Rodrigues, Grebel,
  Pilyugin, Freitas-Lemes, \& Krabbe}]{Dors2015}
Dors, O.~L., Cardaci, M.~V., H{\"a}gele, G.~F., {et~al.} 2015, \mnras, 453,
  4102

\bibitem[{Ferruit {et~al.}(2004)Ferruit, Mundell, Nagar, Emsellem,
  P{\'e}contal, Wilson, \& Schinnerer}]{Ferruit2004}
Ferruit, P., Mundell, C.~G., Nagar, N.~M., {et~al.} 2004, \mnras, 352, 1180

\bibitem[{Fiore {et~al.}(2017)Fiore, Feruglio, Shankar, Bischetti, Bongiorno,
  Brusa, Carniani, Cicone, Duras, Lamastra, Mainieri, Marconi, Menci, Maiolino,
  Piconcelli, Vietri, \& Zappacosta}]{Fiore2017}
Fiore, F., Feruglio, C., Shankar, F., {et~al.} 2017, \aap, 601, A143

\bibitem[{Fitzpatrick(1999)}]{Fitzpatrick1999}
Fitzpatrick, E.~L. 1999, \pasp, 111, 63

\bibitem[{{Garc{\'\i}a-Bernete} {et~al.}(2021){Garc{\'\i}a-Bernete},
  {Alonso-Herrero}, {Garc{\'\i}a-Burillo}, {Pereira-Santaella},
  {Garc{\'\i}a-Lorenzo}, {Carrera}, {Rigopoulou}, {Ramos Almeida}, {Villar
  Mart{\'\i}n}, {Gonz{\'a}lez-Mart{\'\i}n}, {Hicks}, {Labiano}, {Ricci}, \&
  {Mateos}}]{GarciaBernete2021}
{Garc{\'\i}a-Bernete}, I., {Alonso-Herrero}, A., {Garc{\'\i}a-Burillo}, S.,
  {et~al.} 2021, \aap, 645, A21

\bibitem[{{Garc{\'\i}a-Bernete} {et~al.}(2022){Garc{\'\i}a-Bernete},
  {Rigopoulou}, {Alonso-Herrero}, {Donnan}, {Roche}, {Pereira-Santaella},
  {Labiano}, {Peralta de Arriba}, {Izumi}, {Ramos Almeida}, {Shimizu},
  {H{\"o}nig}, {Garc{\'\i}a-Burillo}, {Rosario}, {Ward}, {Bellocchi}, {Hicks},
  {Fuller}, \& {Packham}}]{GarciaBernete2022}
{Garc{\'\i}a-Bernete}, I., {Rigopoulou}, D., {Alonso-Herrero}, A., {et~al.}
  2022, \aap, 666, L5

\bibitem[{{Garc{\'\i}a-Burillo} {et~al.}(2014){Garc{\'\i}a-Burillo}, {Combes},
  {Usero}, {Aalto}, {Krips}, {Viti}, {Alonso-Herrero}, {Hunt}, {Schinnerer},
  {Baker}, {Boone}, {Casasola}, {Colina}, {Costagliola}, {Eckart}, {Fuente},
  {Henkel}, {Labiano}, {Mart{\'\i}n}, {M{\'a}rquez}, {Muller}, {Planesas},
  {Ramos Almeida}, {Spaans}, {Tacconi}, \& {van der Werf}}]{GarciaBurillo2014}
{Garc{\'\i}a-Burillo}, S., {Combes}, F., {Usero}, A., {et~al.} 2014, \aap, 567,
  A125

\bibitem[{{Garc{\'\i}a-Burillo} {et~al.}(2017){Garc{\'\i}a-Burillo}, {Viti},
  {Combes}, {Fuente}, {Usero}, {Hunt}, {Mart{\'\i}n}, {Krips}, {Aalto},
  {Aladro}, {Ramos Almeida}, {Alonso-Herrero}, {Casasola}, {Henkel},
  {Querejeta}, {Neri}, {Costagliola}, {Tacconi}, \& {van der
  Werf}}]{GarciaBurillo2017}
{Garc{\'\i}a-Burillo}, S., {Viti}, S., {Combes}, F., {et~al.} 2017, \aap, 608,
  A56

\bibitem[{Gil~de Paz {et~al.}(2016)Gil~de Paz, Carrasco, Gallego,
  Iglesias-P{\'a}ramo, Cedazo, Garc{\'\i}a~Vargas, Arrillaga, Avil{\'e}s,
  Cardiel, Carrera, Castillo-Morales, Castillo-Dom{\'\i}nguez, de~la
  Cruz~Garc{\'\i}a, Esteban San~Rom{\'a}n, Ferrusca, G{\'o}mez-{\'A}lvarez,
  Izazaga-P{\'e}rez, Lefort, L{\'o}pez-Orozco, Maldonado,
  Mart{\'\i}nez-Delgado, Morales~Dur{\'a}n, Mujica, P{\'a}ez, Pascual,
  P{\'e}rez-Calpena, Picazo, S{\'a}nchez-Penim, S{\'a}nchez-Blanco, Tulloch,
  Vel{\'a}zquez, V{\'\i}lchez, Zamorano, Aguerri, Barrado~y Nav{\'a}scues,
  Bertone, Cava, Cenarro, Ch{\'a}vez, Garc{\'\i}a, Garc{\'\i}a-Rojas, Guichard,
  Gonz{\'a}lez-Delgado, Guzm{\'a}n, Herrero, Hu{\'e}lamo, Hughes,
  Jim{\'e}nez-Vicente, Kehrig, Marino, M{\'a}rquez, Masegosa, Mayya,
  M{\'e}ndez-Abreu, Moll{\'a}, Mu{\~n}oz-Tu{\~n}{\'o}n, Peimbert,
  P{\'e}rez-Gonz{\'a}lez, P{\'e}rez~Montero, Rodr{\'\i}guez,
  Rodr{\'\i}guez-Espinosa, Rodr{\'\i}guez-Merino, Rodr{\'\i}guez-Mu{\~n}oz,
  Rosa-Gonz{\'a}lez, S{\'a}nchez-Almeida, S{\'a}nchez~Contreras,
  S{\'a}nchez-Bl{\'a}zquez, S{\'a}nchez~Moreno, S{\'a}nchez, Sarajedini,
  Silich, Sim{\'o}n-D{\'\i}az, Tenorio-Tagle, Terlevich, Terlevich,
  Torres-Peimbert, Trujillo, Tsamis, \& Vega}]{GildePaz2016}
Gil~de Paz, A., Carrasco, E., Gallego, J., {et~al.} 2016, in Society of
  Photo-Optical Instrumentation Engineers (SPIE) Conference Series, Vol. 9908,
  Ground-based and Airborne Instrumentation for Astronomy VI, ed. C.~J.
  {Evans}, L.~{Simard}, \& H.~{Takami}, 99081K

\bibitem[{{Giordano} {et~al.}(2018){Giordano}, {Racero}, {Norman},
  {Vall{\'e}s}, {Mer{\'\i}n}, {Baines}, {L{\'o}pez-Caniego}, {Mart{\'\i}}, {de
  Teodoro}, {Salgado}, {Sarmiento}, {Guti{\'e}rrez-S{\'a}nchez}, {Prieto},
  {Lorca}, {Alberola}, {Valtchanov}, {de Marchi}, {{\'A}lvarez}, \&
  {Arviset}}]{Giordano2018}
{Giordano}, F., {Racero}, E., {Norman}, H., {et~al.} 2018, Astronomy and
  Computing, 24, 97

\bibitem[{Gonz{\'a}lez~Delgado {et~al.}(2002)Gonz{\'a}lez~Delgado, Arribas,
  P{\'e}rez, \& Heckman}]{GonzalezDelgado2002}
Gonz{\'a}lez~Delgado, R.~M., Arribas, S., P{\'e}rez, E., \& Heckman, T. 2002,
  \apj, 579, 188

\bibitem[{Harris {et~al.}(2020)Harris, Millman, van~der Walt, Gommers,
  Virtanen, Cournapeau, Wieser, Taylor, Berg, Smith, Kern, Picus, Hoyer, van
  Kerkwijk, Brett, Haldane, del R{\'\i}o, Wiebe, Peterson, G{\'e}rard-Marchant,
  Sheppard, Reddy, Weckesser, Abbasi, Gohlke, \& Oliphant}]{Harris2020}
Harris, C.~R., Millman, K.~J., van~der Walt, S.~J., {et~al.} 2020, \nat, 585,
  357

\bibitem[{Hunter(2007)}]{Hunter2007}
Hunter, J.~D. 2007, Computing in Science and Engineering, 9, 90

\bibitem[{{Husemann} {et~al.}(2019){Husemann}, {Scharw{\"a}chter}, {Davis},
  {P{\'e}rez-Torres}, {Smirnova-Pinchukova}, {Tremblay}, {Krumpe}, {Combes},
  {Baum}, {Busch}, {Connor}, {Croom}, {Gaspari}, {Kraft}, {O'Dea}, {Powell},
  {Singha}, \& {Urrutia}}]{Husemann2019}
{Husemann}, B., {Scharw{\"a}chter}, J., {Davis}, T.~A., {et~al.} 2019, \aap,
  627, A53

\bibitem[{{Juneau} {et~al.}(2022){Juneau}, {Goulding}, {Banfield}, {Bianchi},
  {Duc}, {Ho}, {Dopita}, {Scharw{\"a}chter}, {Bauer}, {Groves}, {Alexander},
  {Davies}, {Elbaz}, {Freeland}, {Hampton}, {Kewley}, {Nikutta}, {Shastri},
  {Shu}, {Vogt}, {Wang}, {Wong}, \& {Woo}}]{Juneau2022}
{Juneau}, S., {Goulding}, A.~D., {Banfield}, J., {et~al.} 2022, \apj, 925, 203

\bibitem[{Kakkad {et~al.}(2018)Kakkad, Groves, Dopita, Thomas, Davies,
  Mainieri, Kharb, Scharw{\"a}chter, Hampton, \& Ho}]{Kakkad2018}
Kakkad, D., Groves, B., Dopita, M., {et~al.} 2018, \aap, 618, A6

\bibitem[{Kang \& Woo(2018)}]{Kang2018}
Kang, D. \& Woo, J.-H. 2018, \apj, 864, 124

\bibitem[{Kauffmann {et~al.}(2003)Kauffmann, Heckman, Tremonti, Brinchmann,
  Charlot, White, Ridgway, Brinkmann, Fukugita, Hall, Ivezi{\'c}, Richards, \&
  Schneider}]{Kauffmann2003}
Kauffmann, G., Heckman, T.~M., Tremonti, C., {et~al.} 2003, \mnras, 346, 1055

\bibitem[{Kawamuro {et~al.}(2020)Kawamuro, Izumi, Onishi, Imanishi, Nguyen, \&
  Baba}]{Kawamuro2020}
Kawamuro, T., Izumi, T., Onishi, K., {et~al.} 2020, \apj, 895, 135

\bibitem[{Kewley {et~al.}(2001)Kewley, Dopita, Sutherland, Heisler, \&
  Trevena}]{Kewley2001}
Kewley, L.~J., Dopita, M.~A., Sutherland, R.~S., Heisler, C.~A., \& Trevena, J.
  2001, \apj, 556, 121

\bibitem[{Kewley {et~al.}(2006)Kewley, Groves, Kauffmann, \&
  Heckman}]{Kewley2006}
Kewley, L.~J., Groves, B., Kauffmann, G., \& Heckman, T. 2006, \mnras, 372, 961

\bibitem[{Luridiana {et~al.}(2015)Luridiana, Morisset, \& Shaw}]{Luridiana2015}
Luridiana, V., Morisset, C., \& Shaw, R.~A. 2015, \aap, 573, A42

\bibitem[{Lutz {et~al.}(2020)Lutz, Sturm, Janssen, Veilleux, Aalto, Cicone,
  Contursi, Davies, Feruglio, Fischer, Fluetsch, Garcia-Burillo, Genzel,
  Gonz{\'a}lez-Alfonso, Graci{\'a}-Carpio, Herrera-Camus, Maiolino, Schruba,
  Shimizu, Sternberg, Tacconi, \& Wei{\ss}}]{Lutz2020}
Lutz, D., Sturm, E., Janssen, A., {et~al.} 2020, \aap, 633, A134

\bibitem[{{Maccagni} {et~al.}(2021){Maccagni}, {Serra}, {Gaspari}, {Kleiner},
  {Morokuma-Matsui}, {Oosterloo}, {Onodera}, {Kamphuis}, {Loi}, {Thorat},
  {Ramatsoku}, {Smirnov}, \& {White}}]{Maccagni2021}
{Maccagni}, F.~M., {Serra}, P., {Gaspari}, M., {et~al.} 2021, \aap, 656, A45

\bibitem[{McClintock {et~al.}(1979)McClintock, van Paradijs, Remillard,
  Canizares, Koski, \& V{\'e}ron}]{McClintock1979}
McClintock, J.~E., van Paradijs, J., Remillard, R.~A., {et~al.} 1979, \apj,
  233, 809

\bibitem[{{Meenakshi} {et~al.}(2022){Meenakshi}, {Mukherjee}, {Wagner},
  {Nesvadba}, {Bicknell}, {Morganti}, {Janssen}, {Sutherland}, \&
  {Mandal}}]{Meenakshi2022}
{Meenakshi}, M., {Mukherjee}, D., {Wagner}, A.~Y., {et~al.} 2022, \mnras, 516,
  766

\bibitem[{Mingozzi {et~al.}(2019)Mingozzi, Cresci, Venturi, Marconi, Mannucci,
  Perna, Belfiore, Carniani, Balmaverde, Brusa, Cicone, Feruglio, Gallazzi,
  Mainieri, Maiolino, Nagao, Nardini, Sani, Tozzi, \& Zibetti}]{Mingozzi2019}
Mingozzi, M., Cresci, G., Venturi, G., {et~al.} 2019, \aap, 622, A146

\bibitem[{{Morganti} {et~al.}(2015){Morganti}, {Oosterloo}, {Oonk},
  {Frieswijk}, \& {Tadhunter}}]{Morganti2015}
{Morganti}, R., {Oosterloo}, T., {Oonk}, J.~B.~R., {Frieswijk}, W., \&
  {Tadhunter}, C. 2015, \aap, 580, A1

\bibitem[{{Mukherjee} {et~al.}(2018){Mukherjee}, {Bicknell}, {Wagner},
  {Sutherland}, \& {Silk}}]{Mukherjee2018}
{Mukherjee}, D., {Bicknell}, G.~V., {Wagner}, A.~Y., {Sutherland}, R.~S., \&
  {Silk}, J. 2018, \mnras, 479, 5544

\bibitem[{Nagar {et~al.}(1999)Nagar, Wilson, Mulchaey, \&
  Gallimore}]{Nagar1999}
Nagar, N.~M., Wilson, A.~S., Mulchaey, J.~S., \& Gallimore, J.~F. 1999, \apjs,
  120, 209

\bibitem[{Nelson \& Whittle(1995)}]{Nelson1995}
Nelson, C.~H. \& Whittle, M. 1995, \apjs, 99, 67

\bibitem[{Nesvadba {et~al.}(2008)Nesvadba, Lehnert, De~Breuck, Gilbert, \& van
  Breugel}]{Nesvadba2008}
Nesvadba, N. P.~H., Lehnert, M.~D., De~Breuck, C., Gilbert, A.~M., \& van
  Breugel, W. 2008, \aap, 491, 407

\bibitem[{Pascual {et~al.}(2021)Pascual, Cardiel, Picazo-Sanchez,
  Castillo-Morales, \& Gil De~Paz}]{Pascual2021}
Pascual, S., Cardiel, N., Picazo-Sanchez, P., Castillo-Morales, A., \& Gil
  De~Paz, A. 2021, guaix-ucm/megaradrp: v0.11

\bibitem[{{Pereira-Santaella} {et~al.}(2022){Pereira-Santaella},
  {{\'A}lvarez-M{\'a}rquez}, {Garc{\'\i}a-Bernete}, {Labiano}, {Colina},
  {Alonso-Herrero}, {Bellocchi}, {Garc{\'\i}a-Burillo}, {H{\"o}nig}, {Ramos
  Almeida}, \& {Rosario}}]{PereiraSantaella2022}
{Pereira-Santaella}, M., {{\'A}lvarez-M{\'a}rquez}, J., {Garc{\'\i}a-Bernete},
  I., {et~al.} 2022, \aap, 665, L11

\bibitem[{{Perna} {et~al.}(2020){Perna}, {Arribas}, {Catal{\'a}n-Torrecilla},
  {Colina}, {Bellocchi}, {Fluetsch}, {Maiolino}, {Cazzoli}, {Hern{\'a}n
  Caballero}, {Pereira Santaella}, {Piqueras L{\'o}pez}, \& {Rodr{\'\i}guez del
  Pino}}]{Perna2020}
{Perna}, M., {Arribas}, S., {Catal{\'a}n-Torrecilla}, C., {et~al.} 2020, \aap,
  643, A139

\bibitem[{{Perna} {et~al.}(2017){Perna}, {Lanzuisi}, {Brusa}, {Cresci}, \&
  {Mignoli}}]{Perna2017}
{Perna}, M., {Lanzuisi}, G., {Brusa}, M., {Cresci}, G., \& {Mignoli}, M. 2017,
  \aap, 606, A96

\bibitem[{Pringle {et~al.}(1999)Pringle, Antonucci, Clarke, Kinney, Schmitt, \&
  Ulvestad}]{Pringle1999}
Pringle, J.~E., Antonucci, R. R.~J., Clarke, C.~J., {et~al.} 1999, \apjl, 526,
  L9

\bibitem[{{Rosario} {et~al.}(2019){Rosario}, {Togi}, {Burtscher}, {Davies},
  {Shimizu}, \& {Lutz}}]{Rosario2019}
{Rosario}, D.~J., {Togi}, A., {Burtscher}, L., {et~al.} 2019, \apjl, 875, L8

\bibitem[{Rosario {et~al.}(2010)Rosario, Whittle, Nelson, \&
  Wilson}]{Rosario2010}
Rosario, D.~J., Whittle, M., Nelson, C.~H., \& Wilson, A.~S. 2010, \mnras, 408,
  565

\bibitem[{Schnorr-M{\"u}ller {et~al.}(2014)Schnorr-M{\"u}ller,
  Storchi-Bergmann, Nagar, Robinson, Lena, Riffel, \&
  Couto}]{SchnorrMueller2014}
Schnorr-M{\"u}ller, A., Storchi-Bergmann, T., Nagar, N.~M., {et~al.} 2014,
  \mnras, 437, 1708

\bibitem[{Sharp \& Bland-Hawthorn(2010)}]{Sharp2010}
Sharp, R.~G. \& Bland-Hawthorn, J. 2010, \apj, 711, 818

\bibitem[{{Shimizu} {et~al.}(2019){Shimizu}, {Davies}, {Lutz}, {Burtscher},
  {Lin}, {Baron}, {Davies}, {Genzel}, {Hicks}, {Koss}, {Maciejewski},
  {M{\"u}ller-S{\'a}nchez}, {Orban de Xivry}, {Price}, {Ricci}, {Riffel},
  {Riffel}, {Rosario}, {Schartmann}, {Schnorr-M{\"u}ller}, {Sternberg},
  {Sturm}, {Storchi-Bergmann}, {Tacconi}, \& {Veilleux}}]{Shimizu2019}
{Shimizu}, T.~T., {Davies}, R.~I., {Lutz}, D., {et~al.} 2019, \mnras, 490, 5860

\bibitem[{Silk \& Mamon(2012)}]{Silk2012}
Silk, J. \& Mamon, G.~A. 2012, Research in Astronomy and Astrophysics, 12, 917

\bibitem[{Smirnova-Pinchukova {et~al.}(2019)Smirnova-Pinchukova, Husemann,
  Busch, Appleton, Bethermin, Combes, Croom, Davis, Fischer, Gaspari, Groves,
  Klein, O'Dea, P{\'e}rez-Torres, Scharw{\"a}chter, Singha, Tremblay, \&
  Urrutia}]{SmirnovaPinchukova2019}
Smirnova-Pinchukova, I., Husemann, B., Busch, G., {et~al.} 2019, \aap, 626, L3

\bibitem[{{Smirnova-Pinchukova} {et~al.}(2022){Smirnova-Pinchukova},
  {Husemann}, {Davis}, {Smith}, {Singha}, {Tremblay}, {Klessen}, {Powell},
  {Connor}, {Baum}, {Combes}, {Croom}, {Gaspari}, {Neumann}, {O'Dea},
  {P{\'e}rez-Torres}, {Rosario}, {Rose}, {Scharw{\"a}chter}, \&
  {Winkel}}]{SmirnovaPinchukova2022}
{Smirnova-Pinchukova}, I., {Husemann}, B., {Davis}, T.~A., {et~al.} 2022, \aap,
  659, A125

\bibitem[{{Speranza} {et~al.}(2022){Speranza}, {Ramos Almeida},
  {Acosta-Pulido}, {Riffel}, {Tadhunter}, {Pierce}, {Rodr{\'\i}guez-Ardila},
  {Coloma Puga}, {Brusa}, {Musiimenta}, {Alexander}, {Lapi}, {Shankar}, \&
  {Villforth}}]{Speranza2022}
{Speranza}, G., {Ramos Almeida}, C., {Acosta-Pulido}, J.~A., {et~al.} 2022,
  \aap, 665, A55

\bibitem[{{Talbot} {et~al.}(2022){Talbot}, {Sijacki}, \& {Bourne}}]{Talbot2022}
{Talbot}, R.~Y., {Sijacki}, D., \& {Bourne}, M.~A. 2022, \mnras, 514, 4535

\bibitem[{Thomas {et~al.}(2017)Thomas, Dopita, Shastri, Davies, Hampton,
  Kewley, Banfield, Groves, James, Jin, Juneau, Kharb, Sairam,
  Scharw{\"a}chter, Shalima, Sundar, Sutherland, \& Zaw}]{Thomas2017}
Thomas, A.~D., Dopita, M.~A., Shastri, P., {et~al.} 2017, \apjs, 232, 11

\bibitem[{Ulvestad \& Wilson(1983)}]{Ulvestad1983}
Ulvestad, J.~S. \& Wilson, A.~S. 1983, \apjl, 264, L7

\bibitem[{Ulvestad \& Wilson(1984)}]{Ulvestad1984}
Ulvestad, J.~S. \& Wilson, A.~S. 1984, \apj, 285, 439

\bibitem[{Ulvestad \& Wilson(1989)}]{Ulvestad1989}
Ulvestad, J.~S. \& Wilson, A.~S. 1989, \apj, 343, 659

\bibitem[{Vayner {et~al.}(2017)Vayner, Wright, Murray, Armus, Larkin, \&
  Mieda}]{Vayner2017}
Vayner, A., Wright, S.~A., Murray, N., {et~al.} 2017, \apj, 851, 126

\bibitem[{Veilleux \& Osterbrock(1987)}]{Veilleux1987}
Veilleux, S. \& Osterbrock, D.~E. 1987, \apjs, 63, 295

\bibitem[{Venturi {et~al.}(2021)Venturi, Cresci, Marconi, Mingozzi, Nardini,
  Carniani, Mannucci, Marasco, Maiolino, Perna, Treister, Bland-Hawthorn, \&
  Gallimore}]{Venturi2021}
Venturi, G., Cresci, G., Marconi, A., {et~al.} 2021, \aap, 648, A17

\bibitem[{Virtanen {et~al.}(2020)Virtanen, Gommers, Oliphant, Haberland, Reddy,
  Cournapeau, Burovski, Peterson, Weckesser, Bright, van~der Walt, Brett,
  Wilson, Millman, Mayorov, Nelson, Jones, Kern, Larson, Carey, Polat, Feng,
  Moore, VanderPlas, Laxalde, Perktold, Cimrman, Henriksen, Quintero, Harris,
  Archibald, Ribeiro, Pedregosa, van Mulbregt, \& Contributors}]{Virtanen2020}
Virtanen, P., Gommers, R., Oliphant, T.~E., {et~al.} 2020, Nature Methods, 17,
  261

\bibitem[{{Wagner} \& {Bicknell}(2011)}]{Wagner2011}
{Wagner}, A.~Y. \& {Bicknell}, G.~V. 2011, \apj, 728, 29

\bibitem[{Weinberger {et~al.}(2017)Weinberger, Ehlert, Pfrommer, Pakmor, \&
  Springel}]{Weinberger2017a}
Weinberger, R., Ehlert, K., Pfrommer, C., Pakmor, R., \& Springel, V. 2017,
  \mnras, 470, 4530

\bibitem[{Wilson \& Baldwin(1985)}]{Wilson1985}
Wilson, A.~S. \& Baldwin, J.~A. 1985, \apj, 289, 124

\bibitem[{Wilson \& Ulvestad(1982)}]{Wilson1982}
Wilson, A.~S. \& Ulvestad, J.~S. 1982, \apj, 263, 576

\end{thebibliography}


\begin{appendix}


\clearpage

\section{MEGARA fits with three components for the nuclear region}
\label{app:nucleus}

\begin{figure}[h]
  \centering
  \includegraphics{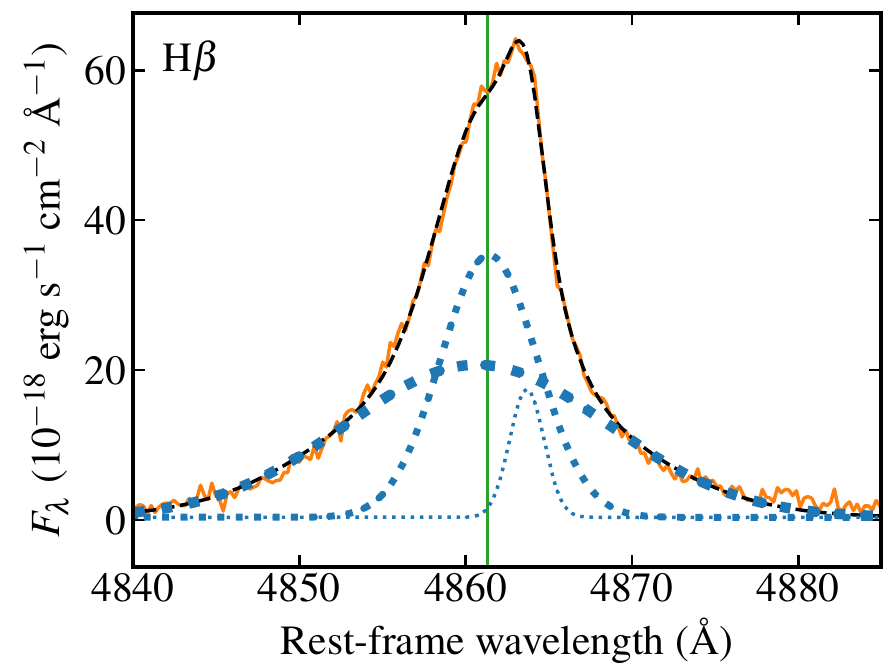}
  \caption{Spectrum of the H$\beta$ line extracted in the nuclear region fitted
    with 3 components. All lines are as in
    Fig.~\ref{fig:spec-center-oiii}.}
  \label{fig:spec-center-hbeta}
\end{figure}

\begin{figure}[h]
  \centering
  \includegraphics{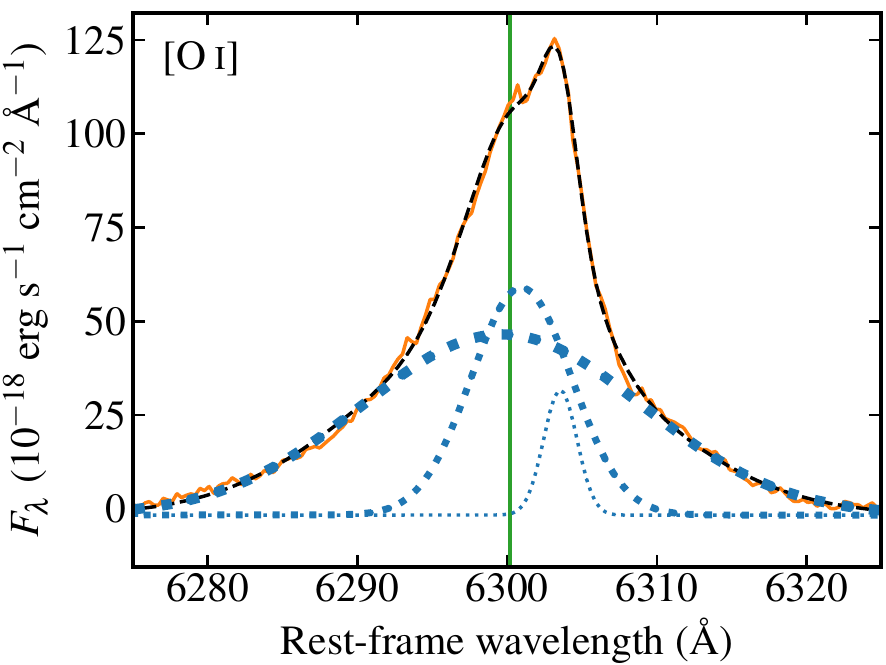}
  \caption{Spectrum of the [\ion{O}{I}]$\lambda$6300 line extracted in the
    nuclear region fitted with 3 components. All lines are as in
    Fig.~\ref{fig:spec-center-oiii}.}
  \label{fig:spec-center-oi}
\end{figure}

\begin{figure}[h]
  \centering
  \includegraphics{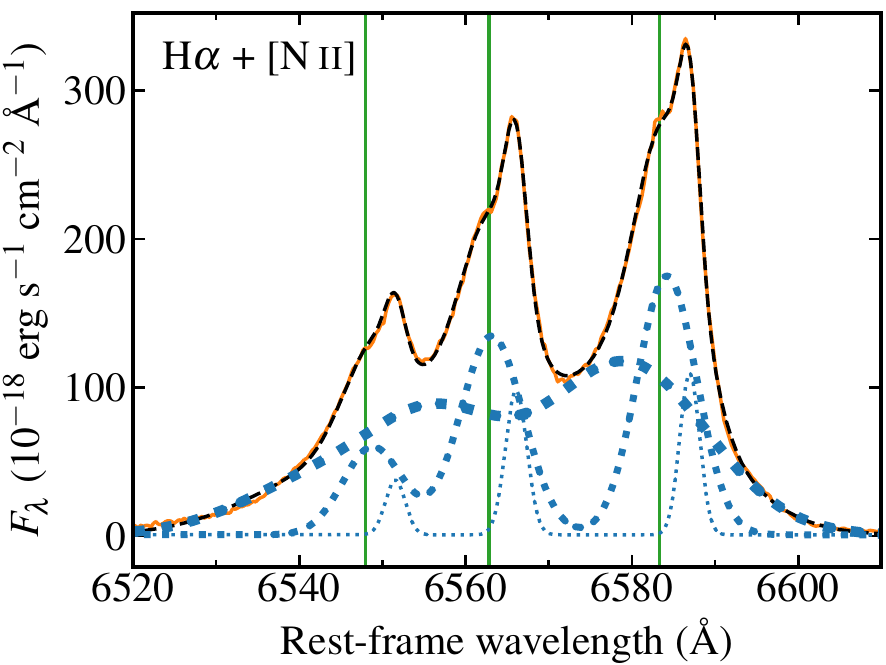}
  \caption{Spectrum of the H$\alpha$ line and the
    [\ion{N}{II}]$\lambda\lambda$6548, 6583 doublet extracted in the nuclear
    region fitted with 3 components. All lines are as in
    Fig.~\ref{fig:spec-center-oiii}.}
  \label{fig:spec-center-halpha-nii}
\end{figure}

\begin{figure}[h]
  \centering
  \includegraphics{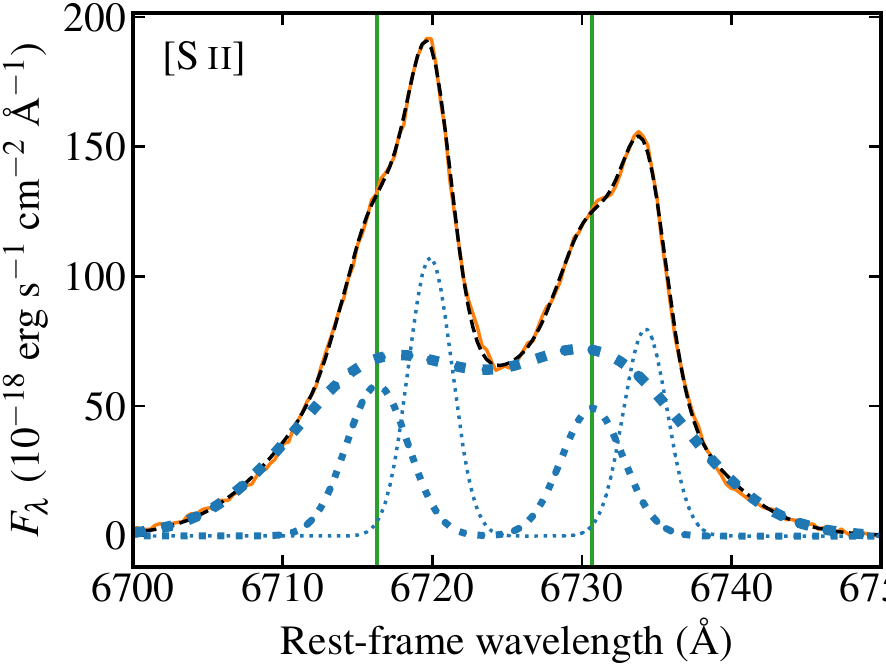}
  \caption{Spectrum of the [\ion{S}{II}] $\lambda\lambda$6716, 6731 doublet
    extracted in the nuclear region fitted with 3 components. All lines are as
    in Fig.~\ref{fig:spec-center-oiii}.}
  \label{fig:spec-center-sii}
\end{figure}

Figs.~\ref{fig:spec-center-hbeta}, \ref{fig:spec-center-oi},
\ref{fig:spec-center-halpha-nii}, and \ref{fig:spec-center-sii} display the
spectra from 3-spaxel square extractions in the nuclear region and their fits
with three components, like Fig.~\ref{fig:spec-center-oiii}
  which shows the
  spectral region around the [\ion{O}{III}]$\lambda\lambda$4959, 5007 doublet in
  the main text. The physical parameters of these fits are provided in
Table~\ref{tab:nucleus}.

\onecolumn


\section{Figures for the emission lines observed with the MEGARA LR-R grating}
\label{app:lr-r}

\begin{figure*}[h]
  \centering
  \includegraphics{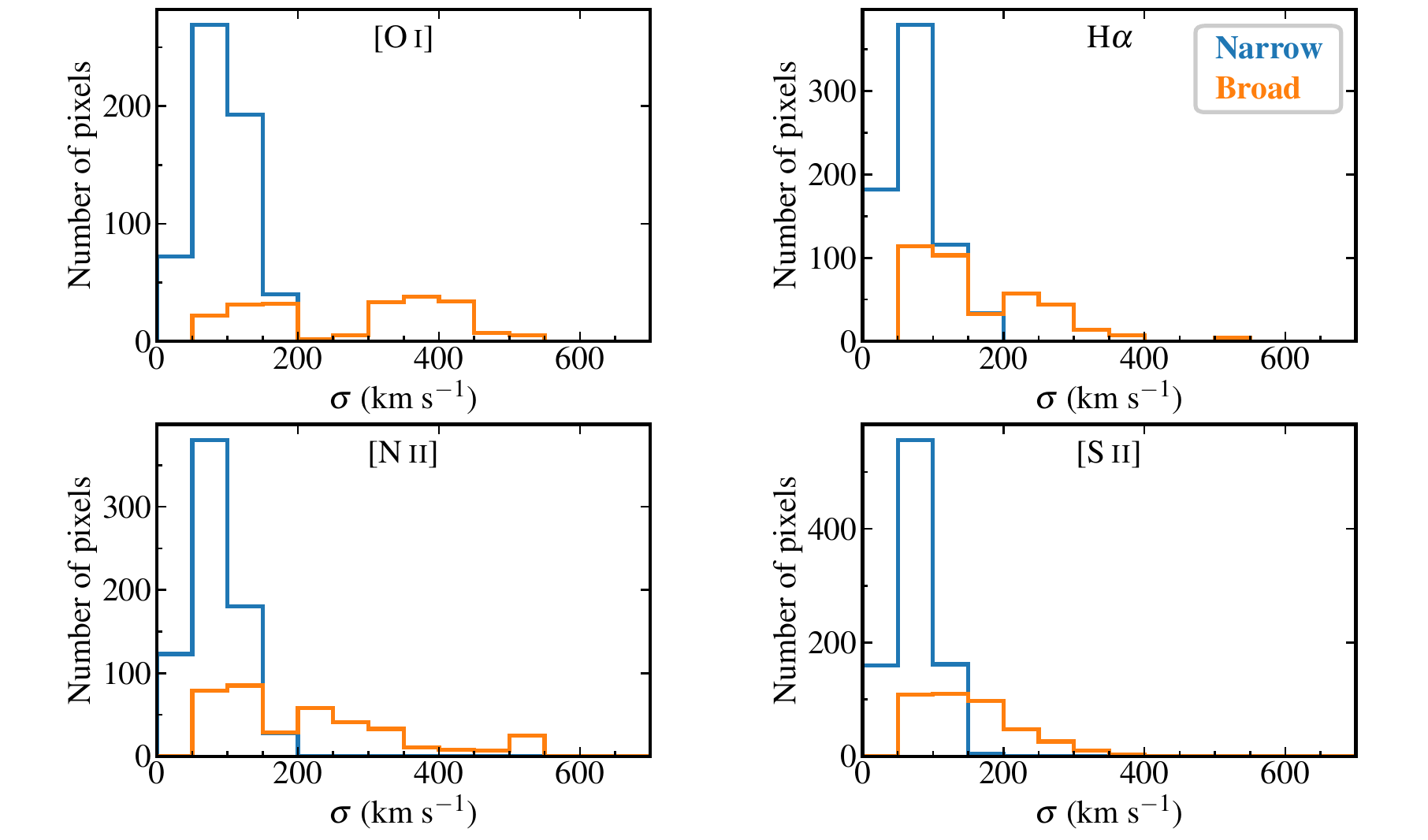}
  \caption{Distributions of the velocity dispersion (corrected for instrumental
    resolution) measured on a spaxel-by-spaxel basis of the fits using a maximum
    of two Gaussian components from lines observed with the MEGARA LR-R grating,
    i.e., [\ion{O}{I}]$\lambda$6300, H$\alpha$,
    [\ion{N}{II}]$\lambda\lambda$6548, 6583 and
    [\ion{S}{II}]$\lambda\lambda$6716, 6731 lines. Same as in
    Fig.~\ref{fig:sigma-hist-blue}.}
  \label{fig:sigma-hist-red}
\end{figure*}

In this Appendix, Figs.~\ref{fig:sigma-hist-red}, \ref{fig:flux-red-a},
\ref{fig:velocity-red}, and \ref{fig:sigma-red} show the same information as
Figs.~\ref{fig:sigma-hist-blue}, \ref{fig:flux-blue}, \ref{fig:velocity-blue},
and \ref{fig:sigma-blue} but for the emission-lines in the MEGARA LR-R grating.

\begin{figure*}[h]
  \includegraphics{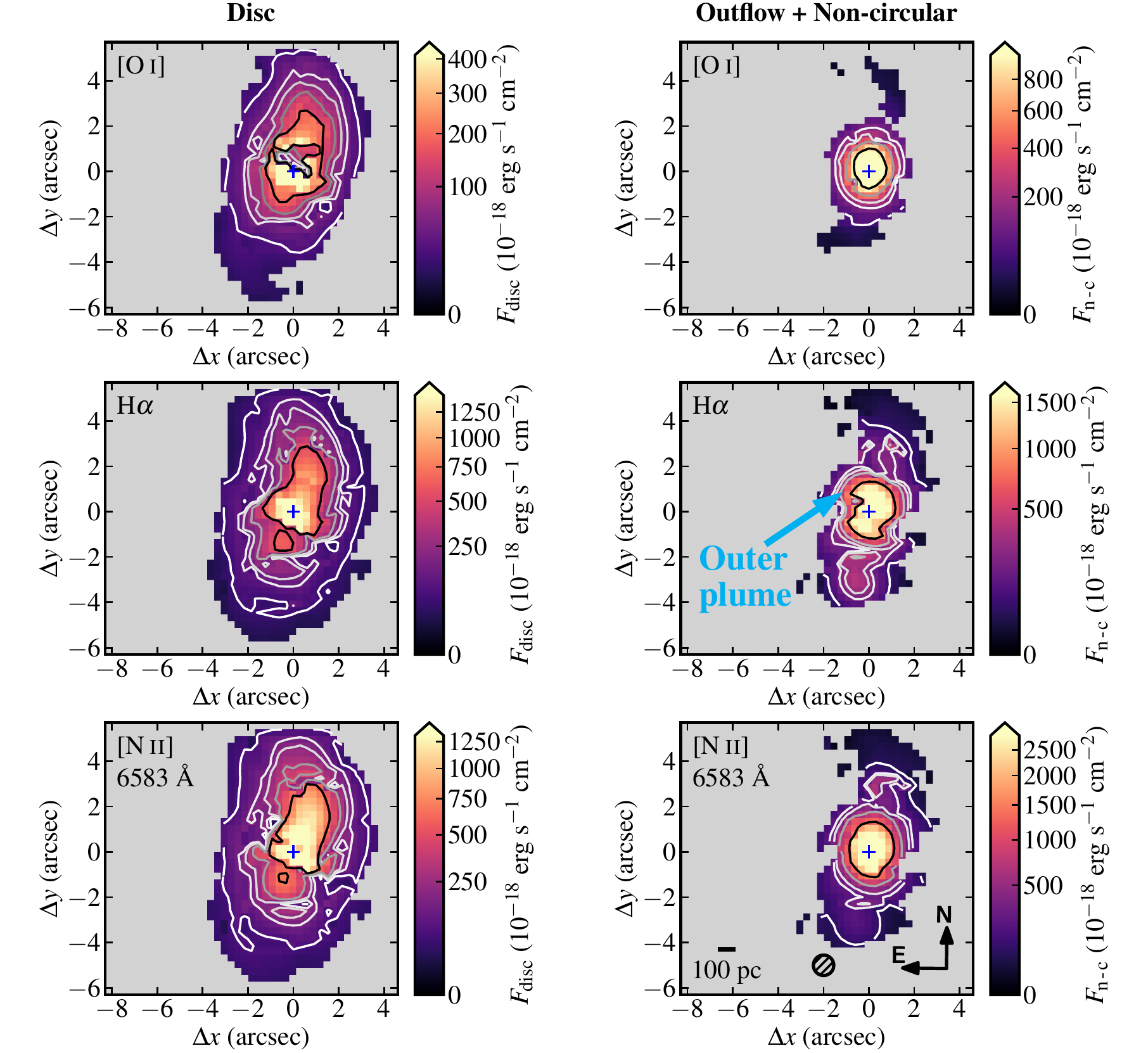}
  \caption{Flux maps of the disc and outflow+non-circular components for lines
    observed with the MEGARA LR-R grating, i.e., [\ion{O}{I}]$\lambda$6300,
    H$\alpha$, [\ion{N}{II}]$\lambda$6583, [\ion{S}{II}]$\lambda$6716 and
    [\ion{S}{II}] $\lambda$6731 lines. Contour definition and symbols are as in
    Fig.~\ref{fig:flux-blue}.}
  \label{fig:flux-red-a}
\end{figure*}

\setcounter{figure}{1}
\begin{figure*}[h]
  \centering
  \includegraphics{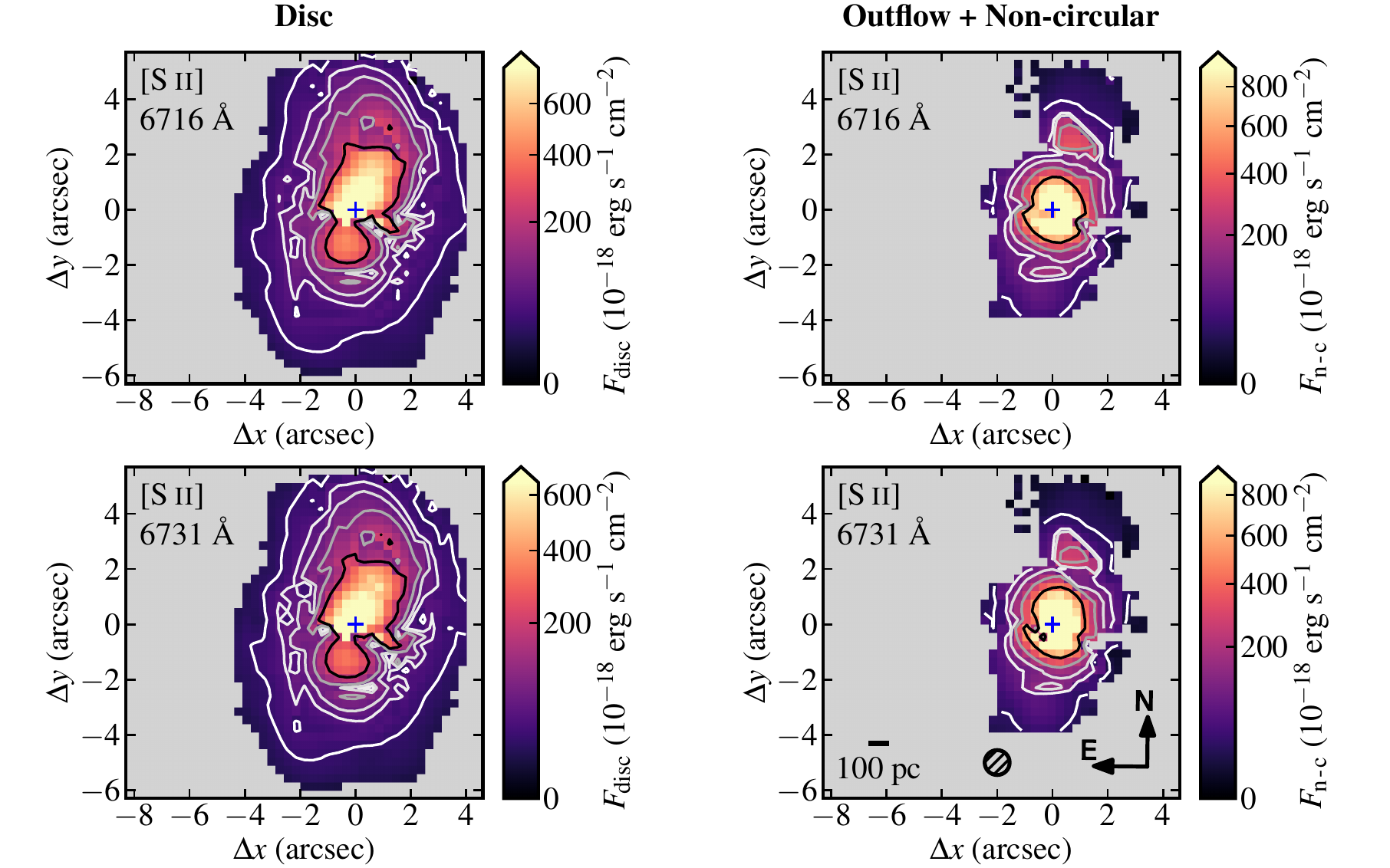}
  \caption{continued.}
  \label{fig:flux-red-b}
\end{figure*}

\begin{figure*}[h]
  \centering
  \includegraphics{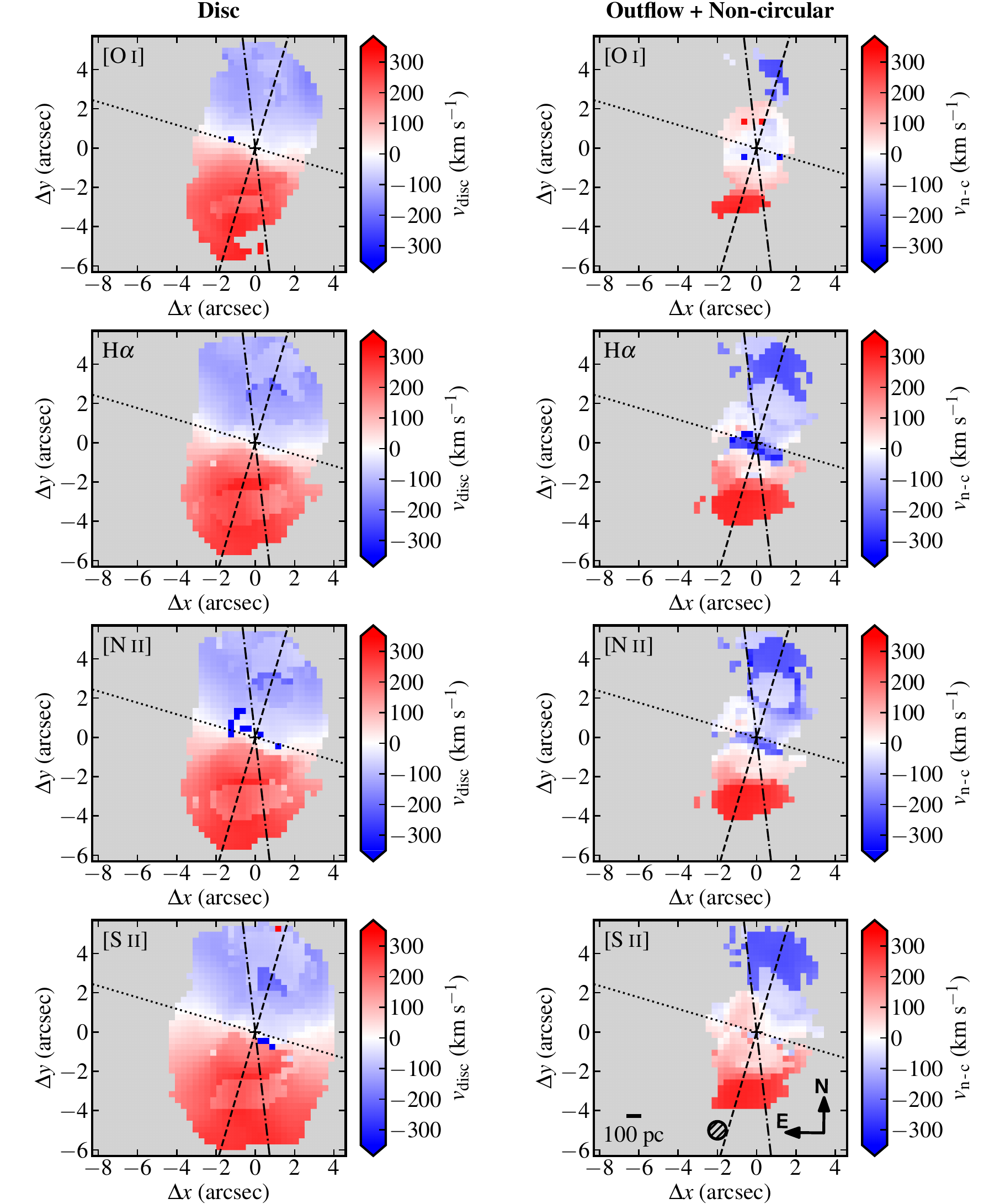}
  \caption{Velocity maps of the disc and outflow+non-circular components for
    lines observed with the MEGARA LR-R grating, i.e.,
    [\ion{O}{I}]$\lambda$6300, H$\alpha$, [\ion{N}{II}]$\lambda\lambda$6548,
    6583 and [\ion{S}{II}]$\lambda\lambda$6716, 6731 lines. Same as in
    Fig.~\ref{fig:velocity-blue}.}
  \label{fig:velocity-red}
\end{figure*}

\begin{figure*}[h]
  \centering
  \includegraphics{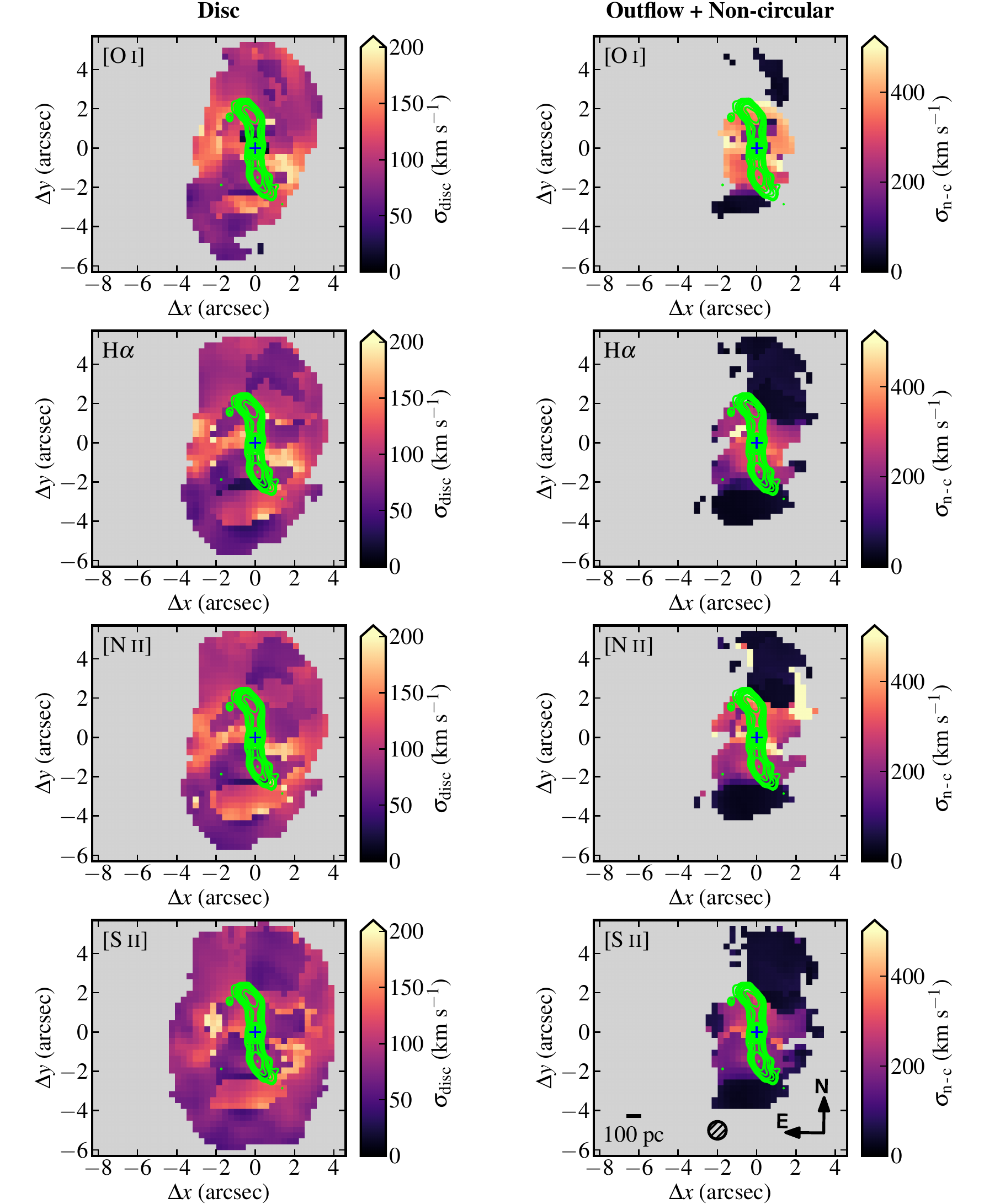}
  \caption{Velocity-dispersion maps of the disc and outflow+non-circular
    components for lines observed with the MEGARA LR-R grating, i.e.,
    [\ion{O}{I}]$\lambda$6300, H$\alpha$, [\ion{N}{II}]$\lambda\lambda$6548,
    6583 and [\ion{S}{II}]$\lambda\lambda$6716, 6731 lines. Same as in
    Fig.~\ref{fig:sigma-blue}.}
  \label{fig:sigma-red}
\end{figure*}


\section{Electron densities from the ionisation parameter method}
\label{app:density-log_u}

\begin{figure*}
  \centering
  \includegraphics{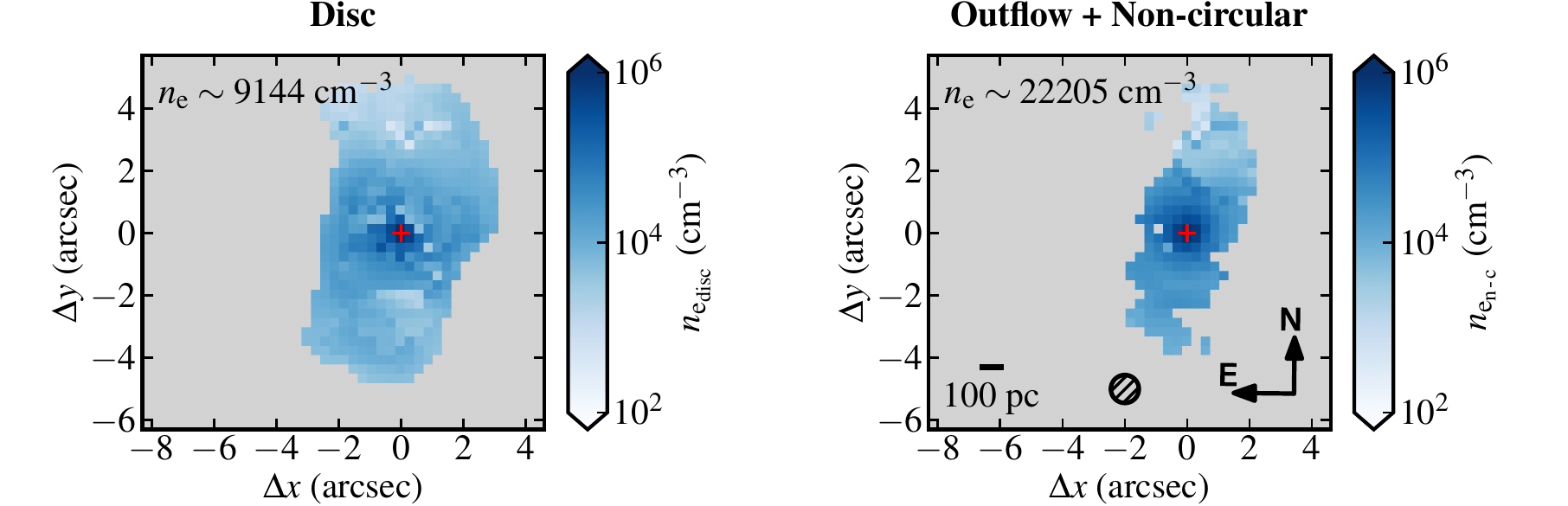}
  \caption{Maps of electron densities for the disc (left) and
    outflow+non-circular (right) components using the ionisation parameter
    method (see text for details).}
  \label{fig:density-log_u}
\end{figure*}

In this Appendix, Fig.~\ref{fig:density-log_u} shows the estimation of
  the electron densities using the ionisation parameter method proposed by
  \citet{Baron2019}. See the main text for details and caveats related to this
  method.

\end{appendix}

\end{document}